\documentclass[11pt]{article}

\usepackage{jheppub}

\usepackage[utf8]{inputenc}
\usepackage{setspace}
\usepackage{verbatim}
\usepackage[usenames,dvipsnames]{xcolor}

\usepackage{amsmath}
\usepackage{latexsym}
\usepackage{tikz}

\usepackage{mathtools}

\usepackage{float}
\usepackage{bm}

\usepackage[normalem]{ulem}

\newcommand{\be}{\begin{equation}}

\newcommand{\Tr}{{\rm Tr }}

\makeatletter
\newcommand{\subalign}[1]{%
  \vcenter{%
    \Let@ \restore@math@cr \default@tag
    \baselineskip\fontdimen10 \scriptfont\tw@
    \advance\baselineskip\fontdimen12 \scriptfont\tw@
    \lineskip\thr@@\fontdimen8 \scriptfont\thr@@
    \lineskiplimit\lineskip
    \ialign{\hfil$\m@th\scriptstyle##$&$\m@th\scriptstyle{}##$\crcr
      #1\crcr
    }%
  }
}
\makeatother

\usepackage{amssymb}
\usepackage{graphicx}
\usepackage{svg}

    \newcommand{\beq}{\begin{equation}}
    \newcommand{\eeq}{\end{equation}}
    \newcommand\beqa{\begin{eqnarray}}
    \newcommand\eeqa{\end{eqnarray}}
\def\<{\left<}
\def\>{\right>}

\newcommand{\eq}[1]{(\ref{#1})}

\def\[{\begin{equation}\begin{aligned}}
\def\]{\end{aligned}\end{equation}}
\newcommand{\tts}[1]{\text{\tiny #1}}
\newcommand{\h}{\mathsf{h}}
\newcommand{\e}{\mathsf{e}}
\newcommand{\f}{\mathsf{f}}

\newcommand{\F}{\mathcal{F}}

\newcommand{\sfa}{\mathsf{a}}
\newcommand{\sfb}{\mathsf{b}}
\newcommand{\E}{\mathbf{E}}
\renewcommand{\S}{\mathbf{S}}
\newcommand{\M}{\mathbf{M}}

\newcommand{\cJ}{\mathcal{J}}

\newcommand{\alg}[1]{\mathfrak{#1}}

\def\<{\big\langle}
\def\>{\big\rangle}

\title{Integrability for the spectrum of Jordanian AdS/CFT}

\emailAdd{sdriezen$\bullet$phys.ethz.ch}
\emailAdd{fedor.levkovich$\bullet$gmail.com}
\emailAdd{amolines$\bullet$phys.ethz.ch}

\author[a]{Sibylle Driezen,}
\author[b]{Fedor Levkovich-Maslyuk,}
\author[a]{Adrien Molines}

\affiliation[a]{Institut f\"ur Theoretische Physik, 
Eidgen\"ossische Technische Hochschule Z\"urich,
Wolfgang-Pauli-Stra\ss e 27, 8093 Z\"urich, Switzerland}

\affiliation[b]{
Centre for Mathematical Science, City St George's, University of London, Northampton Square, EC1V 0HB, London, UK  }

\abstract{
Jordanian deformations offer rare integrable realisations of non-AdS holography, whose solvability methods differ from conventional AdS/CFT examples. Here we study the $\mathfrak{sl}(2,R)$ sector of the Jordanian deformed $AdS_5\times S^5$ string and its weak-coupling spin chain counterpart: the $\mathrm{XXX}_{-1/2}$ model with a non-abelian Jordanian Drinfel'd  twist. While the twist breaks the usual highest-weight structure that underlies conventional Bethe ans\"atze, we show that the complete spectrum remains solvable within the Baxter framework. We argue that the functional form of the $TQ$-relation is unchanged, yet the structure of the $Q$-functions is nontrivially modified. This allows us to obtain analytic expressions at arbitrary spin chain length $J$, which match the deformed string spectrum at the one-loop level and to subleading order in the large-$J$ expansion, despite the severely reduced symmetry. Our results provide nontrivial tests of the Jordanian AdS/CFT correspondence and lay the groundwork for implementing the Separation of Variables program in non-abelian Drinfel'd-twisted models.
}

\begin{document}

\maketitle

\section{Introduction}

Integrable spin chains provide one of the most powerful frameworks for studying exactly solvable quantum systems. 
Their algebraic structure, built from     $R$-matrices satisfying the Yang-Baxter equation, encodes entire families of commuting conserved charges and enables the computation of
complete spectra through the Bethe or Baxter approaches.
Beyond their condensed-matter and mathematical applications, a particularly fruitful role has been their emergence in the planar limit of ${\cal N}=4$ Super-Yang–Mills (SYM) theory, where the one-loop dilatation operator coincides with an $\mathrm{XXX}$-type Hamiltonian, whose long-chain limit reproduces the classical string dynamics on $AdS_5\times S^5$ \cite{Minahan:2002ve,Beisert:2003yb,Beisert:2005fw,Beisert:2010jr}.
This remarkable equivalence has elevated integrable spin chains from solvable toys to structural tools for Quantum Field Theory (QFT), where integrability provides a non-perturbative handle on the (planar) dynamics.

From this perspective, deforming the model while preserving integrability offers a way to probe new QFTs with exact solvability built in. Deformations can   act geometrically, by modifying the target space of the string sigma-model, and algebraically, by modifying the Hopf-algebra symmetry of the spin chain. 
A particularly elegant framework   is given by the Homogeneous Yang-Baxter (HYB) construction \cite{Klimcik:2008eq,Delduc:2013fga,Delduc:2013qra}, where the deformation is induced by a classical $r$-matrix in such a way that a non-local canonical transformation of the sigma-model preserves the classically integrable Lax representation while modifying the string boundary conditions \cite{Frolov:2005dj,Matsumoto:2015jja,Vicedo:2015pna,vanTongeren:2018wek,Borsato:2021fuy} in terms of a classical Drinfel'd twist \cite{vanTongeren:2018wek,Driezen:2025dww}.  
A key  {proposal}, originating from \cite{vanTongeren:2015uha}, is that every HYB deformation is algebraically equivalent to a Drinfel'd twist \cite{drinfeld1983constant} of the {quasi-triangular Hopf algebra} symmetry, which by construction preserves quantum integrability. This connects   string sigma-model deformations with algebraic deformations of the associated spin chains, providing a controlled way to generate new integrable quantum systems whose 
physics can differ drastically from the undeformed model. 

When the twist is generated by commuting  generators\footnote{Such twists, also known as abelian twists or Drinfel'd-Reshetikhin (DR) twists \cite{Reshetikhin:1990ep},  are realised on the sigma-model as TsT-transformations \cite{Osten:2016dvf}. Famous examples include the $\beta$-deformation or Lunin-Maldacena background \cite{Leigh:1995ep,Lunin:2005jy}  and the Schr\"odinger (dipole) deformation \cite{Bergman:2000cw,Alishahiha:2003ru}, cf.~\cite{Beisert:2005if} and \cite{Guica:2017mtd} for their integrability in terms of DR twists.}
which are Cartan, the resulting models   preserve the  Cartan subalgebra of symmetries, and the Bethe ans\"atze continue to apply almost unchanged  \cite{Beisert:2005if,McLoughlin:2006cg,Ahn:2010ws,deLeeuw:2012hp,vanTongeren:2021jhh,Meier:2025tjq}. In contrast, non-Cartan twists, such as the abelian dipole  \cite{Guica:2017mtd} and the non-abelian Jordanian  \cite{Driezen:2025dww} deformation,  modify the algebra in a qualitatively different way,   such that most familiar concepts must be re-examined.\footnote{We note that non-Cartan twists have nontrivial effects on the spectrum only on noncompact algebras. See e.g.~\cite{Kulish_1997,Kulish2009} for the study of the Jordanian twists on the compact $\mathfrak{su}(2)$-invariant spin chain, for which the spectrum remains invariant. } In fact, while they preserve integrability in terms of $R$-matrices and commuting transfer matrices, they render conventional Bethe-ansatz approaches unsuitable to study the full spectrum \cite{Guica:2017mtd,Borsato:2025smn,Driezen:2025dww}, and they also obscure the relation to factorised worldsheet scattering \cite{Borsato:2024sru} (see also \cite{deLeeuw:2025sfs}).  Understanding the exact solvability of these models is therefore of intrinsic algebraic interest. Holographically, they are also  particularly interesting because the corresponding HYB deformations of the $AdS_5\times S^5$ worldsheet   realise Schrödinger geometries \cite{Kawaguchi:2013lba,Matsumoto:2015uja,Guica:2017mtd,Borsato:2022drc},   an important aspect of the non-AdS holography programme. A natural question is thus whether the success of AdS/CFT integrability extends when most of its symmetries are broken by these non-Cartan twists. For the abelian dipole deformation, the nearest-neighbour spin chain spectrum was obtained via the Baxter approach with intricate asymptotics \cite{Guica:2017mtd},  and was shown to match the one-loop worldsheet spectrum \cite{Ouyang:2017yko}.  In this work, we show that such a matching extends nontrivially to the  Jordanian deformations as well, which is particularly surprising given its severely reduced symmetry and non-abelian nature. 

The starting point for our analysis is the non-abelian Jordanian-twisted $\mathfrak{sl}(2,R)$-invariant $\mathrm{XXX}_{-1/2}$ spin chain, following up on the work  \cite{Driezen:2025dww}. There, the algebraic framework underlying its integrability   was constructed and clarified,  and an initial study of the spectral problem for the ground state of the length $J=2$ chain was performed by  direct perturbative diagonalisation. It was also shown that the twist-deformation can be recast as an undeformed chain with twisted boundary conditions determined by a residual root symmetry, which was then used to verify  that the ground state energy of the chain in the classical continuum Landau-Lifshitz limit reproduces  that of  the classical vacuum solution of the Jordanian $AdS_5\times S^5$ string. This provided the first    evidence of the correspondence between the string and spin chain spectrum under non-abelian twists. 

The purpose of the present paper is to analyse the complete spectrum of the Jordanian nearest-neighbour chain and to uncover the full  analytic structure of its integrability. We  re-examine the approach of \cite{Driezen:2025dww} and  show that each undeformed eigenvalue acquires a smooth Jordanian deformation.  Our main outcome is then that, despite the absence of Bethe equations,  the spectrum is fully solvable through the Baxter framework. 

Concretely, we  start from the differential-operator realisation of the transfer matrix and solve its eigenvalue problem  for $J=2$ and arbitrary spin $S$, obtaining closed-form expressions (in $S$) for both the eigenvalues and eigenfunctions as  perturbative series in the deformation parameter.
We  then verify explicitly that the twisted Hamiltonian is diagonal on  the resulting $S=0$ and $S=1$ eigenstates   and compute their corresponding energies. 
Building on these results, we  propose a Baxter relation for the Jordanian-twisted chain whose functional form is the same as that of the undeformed model  (cf.~also \cite{Driezen:2025dww}). The deformation enters only through several fixed coefficients of the transfer matrix. The $Q$-functions cease to be polynomials, and we propose that regularity in the complex plane replaces polynomiality as the condition selecting  physical solutions. Notably, the asymptotics of the $Q$-functions are  simpler than in the abelian dipole case \cite{Guica:2017mtd}. We then show that this framework  provides  the full $J=2$  spectrum very efficiently (including dependence on the spin label $S$) and in perfect agreement with the  nontrivial direct diagonalisation results. This thus gives  compelling evidence for the validity  of the given Baxter framework.  

Finally, we exploit
 the  analytic power of the Baxter relation  to \textit{(i)} extend  the $J=2$ spectrum beyond the perturbative small-deformation  regime using numerical methods, and \textit{(ii)}  
derive the $S=0$ and $S=1$ spectra for arbitrary chain length $J$.
Taking the large-$J$ continuum limit, we find that the leading and subleading corrections reproduce the semiclassical string spectrum, obtained via algebraic curve methods in \cite{Borsato:2022drc}, upon 
an appropriate identification of  residual  charges.
This not only provides an additional nontrivial test of the Baxter approach, but also pushes the Jordanian non-abelian AdS/CFT programme to the one-loop quantum level.

The paper is organised as follows. Section \ref{s:spin-chain} reviews the relevant spectral objects of the Jordanian-twisted $\mathrm{XXX}_{-1/2}$ chain, and diagonalises the transfer matrix and Hamiltonian  for $J=2$ in the small-deformation regime using a Frobenius-based differential analysis. 
Section \ref{sec:Baxter} develops the Baxter framework, which reproduces the $J=2$ spectrum, and uses it to extend the spectrum  numerically to the large-deformation regime and analytically to arbitrary length.   
Section \ref{sec:strings}  sets up the relevant $\mathfrak{sl}(2,R)$ sector of the Jordanian worldsheet model, and compares the spin chain  with the semiclassical string spectrum, showing nontrivial agreement. 
We conclude in section \ref{sec:conclusions} with conclusions  and an outlook.

\section{The $\mathrm{XXX}_{-1/2}$ Jordanian-twisted spin chain} \label{s:spin-chain}

\subsection{Spectral objects}

In this subsection, we present  the key ingredients for the spectral problem of the Jordanian-twisted $\alg{sl}(2,\mathbb{R})$-invariant $\mathrm{XXX}_{-1/2}$ chain. Further details as well as a construction of these objects for Drinfel'd twisted chains with arbitrary twists can be found in   \cite{Driezen:2025dww}.
We begin by recalling the   untwisted XXX chain in its non-compact spin $s=-1/2$ representation \cite{Kulish:1981gi,Tarasov:1983cj,Beisert:2003jj,Beisert:2003yb} (see also \cite{Derkachov:2002tf,Kirch:2004mk}), and then present their Jordanian-twisted counterparts.

\paragraph{$\mathrm{XXX}_{-1/2}$ spin chain.}
We  denote the $\alg{sl}(2,\mathbb{R})$ generators by $X=\{\h,\e,\f\}$, satisfying the following commutation relations
\[ \label{eq:comm-rels-sl2}
[\h , \e] = \e , \qquad [\h,\f]=-\f , \qquad [\e,\f] = -2\h  .
\]
Hence, $\h$ is a Cartan generator while $\e$ and $\f$ are raising and lowering operators respectively. Each physical site of the chain carries  a non-compact $s=-1/2$ representation of $\alg{sl}(2,\mathbb{R})$. In particular, at site $j$ we attach  a highest-weight module $V_j$ in the discrete series, which we parametrise by a continuous parameter $z_j$ on which the $\alg{sl}(2,\mathbb{R})$ generators act with differential operators represented as 
\[
\e_j=-\partial_j,\qquad \h_j=-z_j\partial_j-\frac{1}{2}, \qquad \f_j=-z_j^2\partial_j -z_j , \label{eq:sl_2_explicit_rep}
\]
where we use the notation $\mathbf{S}_j=\{\h_j,\e_j,\f_j\}$ for the generators at each site.
The full Hilbert space of the spin chain is  $\mathcal{H}=\bigotimes_{j=1}^JV_j$, with $J$ its total length, and the total Hamiltonian  reads
\[
H=\frac{\lambda}{4\pi^2}\sum_{j=1}^Jh_{j,j+1},\qquad h_{j,j+1}=\psi(\mathbb{J}_{j,j+1}+1)-\psi(0), \label{eq:sl(2)_hamiltonian}
\] 
with periodic boundary conditions on the nearest-neighbour Hamiltonian densities, i.e.~$h_{J,J+1}=h_{J,1}$. Here $\psi(x)$ is the digamma function and $\mathbb{J}$ is a two-site operator defined implicitly through $\mathbb{J}(\mathbb{J}+1)=\Delta(X\cdot X)=\Delta(\h^2-\tfrac{1}{2}(\e\f+\f\e))$, with the dot product  defined through the Cartan-Killing form, and the coproduct $\Delta:V_j\rightarrow V_{j}\otimes V_{j+1}$ the Hopf-algebraic operation  extending the action of algebra generators to more than one tensor site.  For the untwisted chain, the coproduct is canonical, i.e.~$\Delta(X)=X\otimes 1+1\otimes X$.

While the Hamiltonian \eqref{eq:sl(2)_hamiltonian} is invariant under global $\alg{sl}(2,\mathbb{R})$ transformations (since each $h_{j,j+1}$ depends only on the two-site Casimir), its diagonalisation  is far from straightforward. To tackle the spectral problem one can however employ integrability techniques. These  introduce an auxiliary space $\mathsf{a}$,  typically chosen to be the simplest finite-dimensional representation of $\alg{sl}(2,\mathbb{R})$, i.e.~with generators $X_\sfa$ acting through standard matrix operations as
\[ \label{eq:rep-2D-sl2R}
\h_{\sfa} = \frac{1}{2} \begin{pmatrix}
1 & 0 \\ 0 & -1
\end{pmatrix} , \qquad \e_{\sfa} = \begin{pmatrix}
0 & 1 \\ 0 & 0 
\end{pmatrix} , \qquad \f_{\sfa} = \begin{pmatrix}
0 & 0 \\ -1 & 0
\end{pmatrix}.
\]
The key object is then the auxiliary-physical $R$-matrix (also called Lax-transport matrix)
\begin{equation} \label{eq:Lax_matrix} 
    R_{\mathsf{a}j}(u):=u 1_{\mathsf{a}j}+ 2 i X_{\mathsf{a}}\cdot \textbf{S}_{j}
    =\begin{pmatrix}
      u+i \h_{{j}} & - i\f_{{j}}\\
      i \e_{{j}} & u-i \h_{{j}}
    \end{pmatrix},   
\end{equation}
where
$u \in \mathbb{C}$ is arbitrary and called the spectral parameter. It is used to construct the main spectral objects of integrability: the  monodromy $T_{\mathsf{a}} (u)$ and transfer $\hat{\tau} (u)$ matrices, respectively defined as 
\[ \label{eq:monodromy-undef}
T_{\mathsf{a}} (u) &:= 
R_{\mathsf{a} J} (u) \cdots R_{\mathsf{a} 2} (u) R_{\mathsf{a} 1} (u) , \qquad
\hat{\tau} (u) &:= \Tr_{\mathsf{a}} T_{\mathsf{a}} (u) .
\]
Crucially, the $R$-matrix satisfies the Yang-Baxter Equation (YBE)
\[ \label{eq:YBE}
{ R}_{\sfa \sfb}(u-v) { R}_{\sfa j}(u) { R}_{\sfb k}(v) = { R}_{\sfb k}(v){ R}_{\sfa j}(u){ R}_{\sfa\sfb} (u-v),
\]
with $R_{\sfa \sfb}$ the auxiliary-auxiliary $R$-matrix {obtained from $R_{\sfa j}$ by replacing $\mathbf{S}_j$ by $X_\sfb$}. The YBE guarantees that the transfer matrix commutes  for different values of the spectral parameter
\[
[\hat\tau(u),\hat\tau(v)]=0\quad\forall u,v \in \mathbb{C}\label{eq:transfer_matrix_comm} ,
\]
and since the transfer matrix  by construction is a  polynomial of degree $J$ in $u$,  its expansion generates $J$ 
mutually commuting operators. However, when the auxiliary representation differs from the physical representation, the Hamiltonian itself will  not be part of this spectral expansion. Nevertheless, the commuting operators from the transfer matrix also commute with the physical Hamiltonian, and thus their spectrum provides $J$ conserved quantities.
Therefore, solving the eigenvalue problem of the transfer matrix will  simultaneously yield the  eigenstates of the Hamiltonian as well as the spectrum of conserved charges encoded in $\hat\tau(u)$; for the current model see \cite{Faddeev:1994zg,Derkachov:2002tf}. 

\paragraph{Jordanian twisted chain.} 
We now turn to the Jordanian-twisted $\mathrm{XXX}_{-1/2}$  chain. 
More details on the explicit derivations of the expressions below can  again be found in \cite{Driezen:2025dww}.

The non-abelian Jordanian twist can be represented by\footnote{In fact, there exists many equivalent expressions for the Jordanian twist, which are related by Hopf algebra isomorphisms \cite{Tolstoy:2008zz,Meljanac:2016njp}.} \cite{Gerstenhaber:1992,Ogievetsky:1992ph,Kulish2009}
\[ \label{eq:F-jor}
{\cal F} = \exp \left( \h \otimes \sigma   \right) , \qquad \sigma:= \log (1+\xi \e ) ,
\]
where $\xi \in \mathbb{R}$ will play the role of deformation parameter. One can use it to deform the spin chain by twisting its coproduct and $R$-matrix as
\[
\Delta_{\cal F} (X) := \F \Delta(X) \F^{-1} , \qquad R^{\cal F}(u) := {\cal F}^{\tts{op.}} R(u) \F^{-1},  \label{eq:twisted_coproduct_r_matrix}
\]
with ${\F}^{\tts{op.}}:= P \F P$ and  $P(X \otimes Y) = Y \otimes X$ is a permutation of tensor spaces.
Crucially, the Jordanian twist is a Drinfel'd twist, i.e.~it satisfies the  cocycle condition \cite{drinfeld1983constant}
\[ \label{eq:cocyle-cond}
({\cal F} \otimes 1) (\Delta \otimes 1) ({\cal F}) = (1 \otimes {\cal F}) (1\otimes {\Delta}) ({\cal F}), 
\]
which ensures that the twisted $R$-matrix \eqref{eq:twisted_coproduct_r_matrix} also satisfies the YBE \eqref{eq:YBE}, 
\[
{ R}^\F_{\sfa \sfb}(u-v) { R}^\F_{\sfa j}(u) { R}^\F_{\sfb k}(v) = { R}^\F_{\sfb k}(v){ R}^\F_{\sfa j}(u){ R}^\F_{\sfa\sfb} (u-v) .
\]
As such, deforming the chain by a Drinfel'd twist preserves its integrability. The deformed Hamiltonian is 
\[ \label{eq:Ham-def}
H^{\F} :=\frac{\lambda}{4\pi^2} \sum_{j=1}^J h^{\F}_{j,j+1} , \qquad h^{\F}_{j,j+1} := \F_{j,j+1} h_{j,j+1} \F_{j,j+1}^{-1} ,
\]
with periodic-boundary conditions $h^\F_{J,J+1}=h^\F_{J,1}$, and the deformed monodromy matrix is  
\[
T_{\mathsf{a}}^{\F} (u) :=
R^{\cal F}_{\mathsf{a} J} (u) \cdots R^{\cal F}_{\mathsf{a} 2} (u) R^{\cal F}_{\mathsf{a} 1} (u) .
\]
By construction, $T_{\mathsf{a}}^{\F} (u)$ satisfies a deformed  RTT relation with $R^{\cal F}(u)$,
\[
R^\F_{\sfa \sfb}(u-v)T^\F_\sfa(u)T^\F_\sfb(v)=T^\F_\sfb(v)T^\F_\sfa(u)R^\F_{\sfa \sfb}(u-v).\label{eq:RFTFTF}
\]
However, it turns out that $  \Tr_\sfa T_\sfa^\F(u)$ is very cumbersome to work with as it is not manifestly adapted to the residual global symmetries of the deformed theory. Instead, in \cite{Borsato:2025smn,Driezen:2025dww}, it was shown that a non-local similarity transformation maps the deformed spin chain with periodic boundary conditions and twisted coproduct to an undeformed spin chain with twisted boundary conditions and canonical coproduct. The latter we call the twisted-boundary picture. In particular, the similarity transform is realised by the global intertwiner $\Omega$  which reads (see also \cite{Maillet:1996yy})
\[
\Omega = \F_{12} (\F_{13} \F_{23}) \cdots (\F_{1J} \F_{2J} \cdots \F_{(J-1)J} ) . \label{eq:global_intertwiner}
\]
The Hamiltonian in the twisted-boundary picture then becomes
\[ \label{eq:Ham-twist}
 H^\star := \Omega^{-1} H^{\F} \Omega = \frac{\lambda}{4\pi^2}\left( \sum_{j=1}^{J-1} h_{j,j+1} + B^{-1} h_{J,1} B \right),
\]
with twisted-boundary conditions $h_{J,J+1} = B^{-1} h_{J,1} B$, where $B = \F_{J1}^{-1}\Omega$ \cite{Borsato:2025smn}. Importantly, the nontrivial boundary violates the  full global $\alg{sl}(2,\mathbb{R})$ symmetry. In fact,  $H^\star$ is  only invariant under the action of the raising operator
\[
 [\Delta^{J-1} (\e) , H^{\star}] =0  .
\]
We denote this global operator by $\widehat{M}:=\Delta^{J-1}(\e)=\sum_{j=1}^J\e_j$, and it will prove  invaluable  to work with its eigenstates when studying the spectral problem of the chain.

In addition, the monodromy matrix and its associated transfer matrix in the twisted-boundary picture read \cite{Driezen:2025dww}
\[ \label{eq:jor-twisted-monodromy}
T^{\star}_\mathsf{a} (u):=  \Omega^{-1} T_{\mathsf{a}}^{\F} (u) \Omega = \exp \left( \sigma_{\sfa} \otimes \Delta^{J-1} (\h) \right) T_{\sfa} (u)   \exp \left( -\h_{\sfa} \otimes  \log (1+\xi \widehat{M}) \right),
\]
and
\[
\hat{\tau}^\star(u):={\rm Tr}_\sfa T^\star_\sfa(u) . \label{eq:twisted_transfer_matrix}
\]
Since the global intertwiner $\Omega$ only acts on the physical sites, the twisted monodromy satisfies the same deformed RTT relation as $T^\F(u)$
\[
R^\F_{\sfa \sfb}(u-v)T^\star_\sfa(u)T^\star_\sfb(v)=T^\star_\sfb(v)T^\star_\sfa(u)R^\F_{\sfa \sfb}(u-v) , \label{eq:R*TFTF}
\]
see \cite{Kulish_1997,Borsato:2025smn} for the full expressions of the algebra. 
Noteworthily, the commutation relation for the $C(u)$ element of the monodromy matrix remains undeformed.

At arbitrary length $J$, the twisted transfer matrix  $\hat{\tau}^\star(u)$ was shown in \cite{Driezen:2025dww} to expand  as 
\[
\hat{\tau}^\star(u)=\frac{2+\xi \widehat{M}}{\sqrt{1+\xi \widehat{M}}} u^J + 0 \times u^{J-1} + {\cal O}(u^{J-2}),\label{eq:transfer_matrix_expansion} 
\]
so that the residual symmetry operator $\widehat{M}$ manifestly emerges in its spectral integrability analysis. 
In the following, we  determine the eigenstates of $\hat{\tau}^\star(u)$, which will  be used to diagonalise the twisted Hamiltonian \eqref{eq:Ham-twist}. 
However, since $\hat{\tau}^\star (u)$   acts as a differential operator in $J$ independent variables, its direct diagonalisation  quickly becomes impractical for $J>2$. 
We therefore restrict to the solvable $J=2$ case, which not only allows for an explicit computation but also provides a benchmark for the more powerful Baxter $TQ$ framework of section \ref{sec:Baxter},  which will in turn give access to arbitrary $J$, as well as higher spin and higher-order results.
In particular, here we  determine
the $J=2$ eigenvalue of $\hat{\tau}^\star (u)$ for arbitrary spin $S$ and the corresponding Hamiltonian energies for $S=0,1$, and we will show that these are exactly reproduced within the Baxter   framework.\footnote{This extends the ground-state analysis of \cite{Driezen:2025dww} to higher spin, and, unlike there,   derives the spectrum by diagonalising the Hamiltonian rather than by inserting the  transfer matrix eigenvalue   into the Baxter $TQ$ equation.}

Before starting let us already remark that, although we will label the deformed spectrum by the  spin number $S$ of the undeformed eigenstates, the deformed eigenstates are not eigenstates of the quadratic Casimir. Rather, we use $S$ only as a book-keeping index labelling deformations of undeformed states.

\subsection{$J=2$ eigenvalue problem} \label{ss:J=2-tau}

The first step in computing the $J=2$ spectrum is the explicit diagonalisation of the  Jordanian transfer matrix \eqref{eq:twisted_transfer_matrix}, i.e.~we seek eigenfunctions $\Psi^\star(z_1,z_2)$ satisfying $\hat{\tau}^\star(u)\Psi^\star(z_1,z_2)={\tau}^\star(u)\Psi^\star(z_1,z_2)$ with eigenvalue $\tau^\star(u)$. 
Exploiting the chain's residual symmetry, we  look for simultaneous eigenfunctions of $\widehat{M}$ and $\hat{\tau}^\star(u)$ as   in \cite{Driezen:2025dww},   which reduces the partial differential problem in two variables to an effective single-variable Ordinary Differential Equation (ODE). With the $\alg{sl}(2,\mathbb{R})$ representation \eqref{eq:sl_2_explicit_rep} and  $\widehat{M}=\e_1+\e_2$, this means that we  start with
\[
\Psi^\star(z_1,z_2)= e^{-M x} g^\star(z), \quad \text{with}\quad x:=\frac{z_1+z_2}{2},\quad \text{and} \quad z:=z_2-z_1,\label{eq:J=2_M_eigenstates}
\]
with $M$ the eigenvalue of $\widehat{M}$ and $g^\star(z)$ a yet unknown function of the relative coordinate $z$. Such eigenfunctions were also used in the spectral analysis of the dipole-twisted $\alg{sl}(2,\mathbb{R})$ spin chain \cite{Guica:2017mtd}, and they can be interpreted as Barut-Girardello coherent states in the (global) tensor-product representation, see e.g.~\cite{Barut:1970qf}.

The $J=2$ transfer matrix eigenvalue then takes the form
\[
\tau^\star(u)=\frac{2+ \xi M}{\sqrt{1+ \xi M}} u^2 +\tau^\star_0,
\]
where $\tau^\star_0$ is determined by the ${\cal O}(u^0)$  term of $\hat{\tau}^\star(u)\Psi^\star(z_1,z_2)={\tau}^\star(u)\Psi^\star(z_1,z_2)$, which gives an ODE for $g^\star(z)$,
\begin{equation} 
    p_1(z) g^\star(z) + p_2(z) g^{\star'}(z) + p_3(z) g^{\star''}(z) - 8 \xi z^2 g^{\star'''}(z)=0, \label{eq:J=2_deformed_EV_eq_u^0}
\end{equation}
which is of third order and has coefficients 
\[ \label{eq:p-s}
    p_1(z)&:=4 + M (2 \xi + M z (2 \xi - z (2 + \xi M))) + 
 8 \sqrt{1 + \xi M} \tau^\star_0,\\
    p_2(z)&:=2 (8 z  -4\xi + \xi M z (4 + M z) ),\\
    p_3(z)&:=4 z (2 z - 6 \xi + \xi M z  ). 
\]
This equation coincides with eq.~(5.17) of \cite{Driezen:2025dww}. 
There, the solution was obtained by the Frobenius method around the singular point at $z=0$, and imposing regularity of the result in the undeformed limit. This yielded a solution for the deformed ground state, while excited states appeared absent in this method. 
Here, we revisit this equation and show that first performing a perturbative expansion in the deformation parameter $\xi$, rather than $z=0$, naturally reveals the deformed spectrum including the excited states. We comment further on the relationship between these two  approaches at the end of this subsection.

In the undeformed $\xi=0$ case, the ODE \eqref{eq:J=2_deformed_EV_eq_u^0} reduces to a second-order equation whose regular solutions around $z=0$ are
\[\label{eq:J=2_undef}
g(z)=\mathrm{SJ}\left(S;-\frac{i}{2} M z\right),\qquad \tau_0=-\frac{1}{2} -S(S+1),
\]
with $\mathrm{SJ}$  the spherical Bessel function of the first kind and $S\in\mathbb{N}$  the magnon number (the number of spin excitations of the state). To incorporate the Jordanian deformation, we expand both the wavefunction and its eigenvalue   in the parameter $\xi M$ around their undeformed solutions\footnote{Given the asymptotics of $\tau^\star (u)$ as well as the classical energy of the corresponding string solution, the spectrum of the Jordanian chain is expected to depend on the deformation parameter only through the combination $\xi M$, see also \cite{Driezen:2025dww}. The deformation of the spectrum is thus nontrivial only for states with a nonzero $M$ charge. \label{f:xi-and-M} }
\[
g^\star(z)=g(z)+\sum_{l=1} (\xi M)^l g_l(z), \qquad \tau_0^\star=\tau_0+\sum_{l=1}(\xi M)^l \tau_l . \label{eq:J=2_perturbative_ansatz} 
\]
Substituting these ans{\"a}tze
into \eqref{eq:J=2_deformed_EV_eq_u^0} and expanding order by order in $\xi M$ yields a hierarchy of \textit{second-order} ODEs for the $g_l(z)$.\footnote{Expanding in $\xi$ effectively restores the order of the undeformed problem:  
since the $\xi=0$ equation is of second order, only two linearly independent solutions will admit a regular power-series expansion in $\xi$.}
At each order the equation has a singular point at $z=0$, which is of the ``regular'' type that allows a Frobenius (regular power-series) expansion in $z$
\[
g_l(z)=z^{k_l} \sum_{j=0} g_{l,j} z^j , \label{eq:J=2_frobenius_ansatz}
\]
with $k_l$ the characteristic exponent, which we demand to be positive to ensure regularity at $z=0$. Plugging this ansatz into the hierarchy of ODEs, we observe that $k_l=S-l$ for generic $S$. Now, note that at each order in $\xi$ a physically irrelevant integration constant will appear, reflecting the freedom to multiply the eigenstates by an overall, possibly $\xi$-dependent, normalisation. We fix this freedom by observing that the 
the leading behaviour of  the undeformed solution $g(z)$ around $z= 0$ is always
\[
g(z) \sim z^S , \qquad z \sim 0 .
\]
In other words, the ${\cal O}(z^S)$ term of $g(z)$  is always nonvanishing, and one can use this fact to demand that the ${\cal O}(z^S)$ term of $g^\star (z)$ remains undeformed.
Explicitly, given the ansatz \eqref{eq:J=2_perturbative_ansatz} and $k_l=S-l$, the coefficient of $z^S$ in $g^\star(z)$ reads
\[
z^S\left(1+\sum_{l=1}^\infty (\xi M)^l g_{l,l}\right),\label{eq:normalisation_prescription}
\]
and we thus fix $g_{l,l}=0$. 
This prescription simply removes  the unphysical overall scaling freedom in the perturbative expansion, and we have explicitly verified that the transfer matrix eigenvalue is independent of this choice.\footnote{An analogous normalisation condition will appear in section \ref{sec:Baxter},  when fixing the normalisation of the $Q$-functions.}

Solving the equations order by order in $z$ and  in $\xi M$, we then obtain
for the transfer matrix eigenvalue
\[
\tau^\star_0={}&-\frac{1}{2}-S(S+1)- \frac{2 S^2 (S+1)^2-1}{64 S (S+1)-48} \ \xi^2M^2\left(1-\xi M\right) 
+{\cal O}(\xi^4 M^4),            \label{eq:J=2_eigenvalue_all_S}
\]
and for the corresponding eigenfunctions: At  ${\cal O}(\xi M)$ 
\[
g_1(z)= (&-i z M)^{S-1}\sqrt{\pi}\Bigg(-\frac{ 4^{-S-1} S^2}{ \Gamma
   \left(S+\frac{3}{2}\right)}
   -\frac{ 2^{-2 (S+3)} \left(S^2-2\right) z^2M^2  }{\Gamma \left(S+\frac{5}{2}\right)}\\
  & -\frac{  2^{-2 S-11} \left(S^2-4\right) z^4M^4 }{\Gamma\left(S+\frac{7}{2}\right)}
   -\frac{  2^{-2 S-15} \left(S^2-6\right) z^6 M^6 }{3 \Gamma \left(S+\frac{9}{2}\right)}+{\cal O}(z^7)\Bigg),
\]
and at  ${\cal O}(\xi^2 M^2)$ 
\[
g_2(z)= (-izM)^{S-2} \sqrt{\pi}&\Bigg(
    \frac{(2^{ - 2 S-3}   ( S-1)^3 S^2)}{
     ( 2 S-1) \Gamma\left(S+\frac{3}{2}\right)} +\frac{(
    2^{ - 2 S-3}   S^2 z M)}{\Gamma\left(S+\frac{3}{2}\right))} \\
   & + \frac{(
    2^{ - 2 S-7}    ( S^2-2) z^3M^3)}{
  \Gamma\left(S+\frac{5}{2}\right)} +{\cal O}(z^4)\Bigg) . 
\]
In practice, we have computed the eigenvalues $\tau^\star_0$ and eigenfunctions $g^\star (z)$ for arbitrary $S$ up to ${\cal O}(\xi^4M^4)$ included, and the Frobenius expansion of each $g_l(z)$ up to ${\cal O}(z^6)$ included. However, above we only present the truncated solutions for brevity.  In principle, the expansion can be extended to arbitrary order. For the deformations of   the lowest excitations $S=0,1$, we  obtained compact closed-form expressions in $z$, which are presented in appendix \ref{app:J=2_eigenstates}.

\paragraph{Some remarks.}
Setting $S=0$  recovers the ground state  and corresponding eigenvalue  computed in \cite{Driezen:2025dww}, cf.~eqs.~(5.28-5.30) therein. For future convenience, we recall the latter here
\[
 \tau^\star(u)= \frac{2+ \xi M}{\sqrt{1+ \xi M}} u^2   -\frac{1}{2}-\frac{\xi ^2 M^2}{48}+\frac{\xi ^3 M^3}{48}-\frac{667 \xi ^4 M^4}{34560}+\frac{307 \xi ^5 M^5}{17280}+{\cal O}(\xi^6 M^6) 
   .\label{eq:J=2_deformed_eigenvalue}
\] 
The pattern of the expansion of excited states is also noteworthy, 
with gamma functions of the form $\Gamma(S+\tfrac{2n+1}{2})$ repeatedly appearing, hinting  at possible closed-forms for both  eigenfunctions and  eigenvalues. 
Furthermore,  the analysis clearly shows that each undeformed excited eigenstate admits a smoothly deformed counterpart,  with a corresponding deformation of $\tau^\star(u)$, demonstrating that the full undeformed spectrum deforms coherently under the Jordanian twist. Finally,  it is also instructive to compare the present 
approach, based on a perturbative expansion in $\xi$,  with that of \cite{Driezen:2025dww}, where regularity in $z$ was imposed first.
The two approaches appear to be related by 
a nontrivial treatment of integration constants:  allowing these constants to depend on $\xi$   effectively interchanges the order of expansions and allows 
to find excited states above the deformed vacuum. 
Indeed, taking the leading integration constant in eq.~(5.22) of \cite{Driezen:2025dww} to be $g_0=\xi^S$ for $S=0,1,2$ rather than $g_0=1$ prior to demanding regularity at $\xi=0$, reproduces  precisely the $S=0,1,2$ deformed eigenstates and eigenvalues presented in this section.
However, extending this treatment to arbitrary $S$ remains technically cumbersome, as it would require solving  the recurrence relation eq.~(5.23) of \cite{Driezen:2025dww} in closed form such that   the choice of integration constants can be tracked.
By contrast, expanding directly in $\xi$ as done above provides a  more efficient and systematic solution scheme  for all $S$.

\subsection{$J=2$ energy spectrum }
We now use the eigenstates obtained in the previous subsection to extract  the spectrum of the $J=2$ spin chain by directly evaluating the twisted Hamiltonian $H^\star$ \eqref{eq:Ham-twist}. 
To this end, it is convenient to employ  the so-called light-ray representation of the Hamiltonian \cite{Belitsky:2004sc}, in which the local two-site operator $h_{12}$ in \eqref{eq:sl(2)_hamiltonian} acts on arbitrary wavefunctions $\Psi(z_1,z_2)$ as 
\[
h_{12}\Psi(z_1,z_2)=\frac{1}{2} \int_0^1\frac{du}{u}  \left(2 \Psi \left(z_1,z_2\right)-\Psi \left(z_1,z_2-uz_{21}\right)-\Psi \left(z_1+u z_{21},z_2\right)\right),\label{eq:ham_int_rep}
\]
with $z_{21}=z_2-z_1$. This integral representation provides a direct way to evaluate the twisted Hamiltonian  \eqref{eq:Ham-twist} on the eigenstates of the transfer matrix. For $J=2$, $H^\star$ explicitly reads 
\[
H^\star=\frac{\lambda}{4\pi^2}\left(h_{12}+\F_{12}^{-1}\F_{21}h_{12}\F_{21}^{-1}\F_{12}\right).\label{eq:J=2_twisted_ham}
\]
Given the structure of the eigenstates and the expectation that the resulting energy should  depend only on $\xi M$, we can simplify its evaluation  by working in an expansion in $M$.
This approach has the advantage that, at each fixed order in $M$, only finitely many coefficients $g_{l,j}$ contribute.
However, even in the undeformed limit the leading behaviour of the eigenfunctions is $M^S$ for arbitrary $S$, while the undeformed energy starts at ${\cal O}(M^0)$. Consequently,  the ${\cal O}(M^n)$ energy correction  receives contributions from the $(n+S)$-th order in $M$ of the wavefunction.
As a result,  this calculation becomes increasingly technical for generic $S$, and we  therefore restrict our explicit evaluation to the lowest excitations, $S=0,1$. 

Explicitly for $S=0$, the eigenstate reads  
\[
\Psi^\star_{S=0}(z_1,z_2)=1-\frac{1}{2} M \left(z_1+z_2\right)+\frac{1}{24} M^2 \left(\xi  z_{21}+4 z_1^2+4 z_2 z_1+4 z_2^2\right)+{\cal O}(M^3) ,\label{eq:J=2_eigenstate_expand_in_M}
\]
which is exact in $\xi$ and in the coordinates $z_1,z_2$ up to order $M^2$. The twist operator $\F_{12}$ can likewise be expanded in $\xi$ to the same order as the eigenstate in $M$, since the $n$-th order in $\xi$ acting on $\Psi^\star$ can at most generate $n$ additional powers of $M$. For instance, at order $\xi^2$ we have
\[
\F_{12} = 1+ \xi \h_1 \e_2 + \frac{\xi^2}{2} (\h^2_1  \e^2_2- \h_1  \e^2_2) + {\cal O}(\xi^3) .
\]
Although not shown here for brevity, we have worked with $\Psi^\star_{S=0}(z_1,z_2)$ and $\F_{12}$ up to orders ${\cal O}(M^6)$ and ${\cal O}(\xi^6)$ respectively. 
Applying the twisted Hamiltonian to $\Psi^\star_{S=0}(z_1,z_2)$ shows that it is indeed a simultaneous eigenfunction,  and the corresponding energy eigenvalue is
\[
E_{S=0}=\frac{\lambda}{4\pi^2}\left(\frac{\xi^2 M^2}{24} - \frac{\xi^3 M^3}{24}+ \frac{131 \xi^4 M^4}{3456}-\frac{59\xi^5 M^5}{1728}+\frac{38441 \xi^6M^6}{1244160}+{\cal O}(M^7)\right).\label{eq:J=2_energy_check}
\]
This reproduces exactly the ground state energy  computed in \cite{Driezen:2025dww} from the Baxter $TQ$ relation (cf.~eq.~(5.41) therein). 
Although not a formal proof, this constitutes a strong and nontrivial check of the assumption made in \cite{Driezen:2025dww} concerning the validity of the Baxter $TQ$ relation and of the associated energy formula. It thereby provides compelling  support for employing the $TQ$ framework later in section \ref{sec:Baxter}.

Proceeding analogously for the $S=1$ eigenstate, we have
\[
\Psi^\star_{S=1}(z_1,z_2)=-\frac{1}{12} i M \left(2 z_{21}-\xi \right)+\frac{1}{24} i M^2 \left(-\xi ^2 +( z_1+z_2) \left(z_{21}-\xi\right)\right) +{\cal O}(M^3),
\]
which leads to the energy        
\[
E_{S=1}=\frac{\lambda}{4\pi^2}\left(2 h(1) + \frac{\xi^2 M^2 }{40}-\frac{\xi^3 M^3}{40}+{\cal O}(\xi^4 M^4)\right),\label{eq:J=2_energyy_S=1}
\]
with $h(1)$ the first harmonic number.
This result  will likewise be matched to a $TQ$-based computation, once again providing strong support of its usage. 

\section{The spectrum from the Baxter equation}
\label{sec:Baxter}

In this section we present the Baxter equation for the Jordanian deformed spin chain, and show how it can be used to compute its spectrum. In particular, we will reproduce key  results from the previous section in a considerably more efficient way, as well as obtain many new results for the spectrum (including at arbitrary length $J$ and spin $S$).

\subsection{Comments on the deformation structure and usage of Bethe ans\"atze}

Despite the fact that the Jordanian deformed $\mathrm{XXX}_{-1/2}$ spin chain is constructed from the start as an integrable model, utilising this integrability to actually compute key observables like the spectrum is a nontrivial task (as illustrated in the previous section). Before introducing the Baxter approach, let us  outline why the most standard integrability method,  the Bethe ansatz, breaks down in this setting, as already noted in \cite{Driezen:2025dww}.

The difficulty originates from  the interplay between the Jordanian twist structure 
and the requirements of the Bethe ansatz.
The twisted monodromy matrix   \eqref{eq:jor-twisted-monodromy} contains the global twist 
operator $\beta_{\sfa}=\exp \left( -\h_{\sfa} \otimes \log (1+\xi \widehat{M}) \right)$, 
which acts nontrivially in the physical space with the residual symmetry generators $\widehat{M}$. In fact, to probe 
the deformed spectrum, one needs states with nontrivial  charge $M$ under this symmetry; however, such states cannot serve as  conventional  pseudovacuum reference states  on top of which one would normally apply e.g.~the algebraic Bethe ansatz and build magnon excitations.\footnote{See e.g. \cite{ slavnov2019algebraicbetheansatz} for a pedagogical discussion of the usual Bethe ansatz including the simplest twisted case.} 
This feature is quite similar to what happens in the dipole-deformed model from \cite{Guica:2017mtd}. 

Nevertheless, one might hope to remedy this by diagonalising the twists through a similarity transformation. In fact, in contrast to \cite{Guica:2017mtd}, here one has the global twist $\beta_{\sfa}$, which is diagonal in the auxiliary space, on top of the Jordan block-like twist of \cite{Guica:2017mtd}.
One might then naively expect this contribution to render the overall twist diagonalisable. Indeed, if the twists were merely numerical $2\times 2$ matrices, one could find a global change of basis to diagonalise them,  and  the new pseudovacuum would be straightforward to construct  (see e.g.~\cite{Guica:2017mtd,Gromov:2016itr,Jiang:2015lda}). However, in our case the twist matrices are in fact themselves operator-valued, built from non-commuting $\mathfrak{sl}(2,R)$ generators acting on the physical space. These operators cannot be simultaneously diagonalised, and consequently no simple diagonal frame exists. As a result, the usual Bethe ansatz simply cannot be used for our model.

It is also worth noting  that these non-Cartan twists have nontrivial effects on the spectrum only for noncompact algebras, and that Bethe ansatz constructions have been developed for related Jordanian models in more restrictive setups where the spectrum remains undeformed. In particular, in the compact $s=+1/2$ chain, the global $\widehat{M}$ operator is a strictly nilpotent matrix, and therefore there simply do not exist   $\widehat{M}$-eigenstates with $M\neq 0$. While one can initiate the usual Bethe ansatz methods on reference states with $M=0$, the resulting spectrum will be undeformed (see e.g.~\cite{Kulish_1997,Kulish2009} and references therein).
Similarly, in the non-compact $s=-1/2$ chain also studied here, the analysis of \cite{Borsato:2025smn} restricts to subspaces with a maximum value under $\Delta^{J-1}(\h)$, on which effectively $\widehat{M}$ is again nilpotent  and the Hamiltonian is non-diagonalisable. While this again excludes states with nontrivial $M$,  
one can apply Bethe-ans\"atze treatments (on the same pseudovacuum state as in the undeformed model), but  the resulting spectrum  remains undeformed. 

In contrast, in the setup we consider here, with the complete highest-weight discrete  module of the $s=-1/2$ representation of $\mathfrak{sl}(2,R)$, the operator $\widehat{M}$ is no longer nilpotent and genuine eigenstates with $M\neq 0$ exist. The spectrum will be  nontrivially deformed, but
 now the issue is that  eigenstates of $\widehat M$ are not proper reference states for standard Bethe ans\"atze. We therefore turn to the alternative Baxter approach, described in detail below.

\subsection{Baxter equation and energy formula}

A powerful way to solve models without a proper pseudovacuum state is often provided by the quantum separation of variables (SoV) approach \cite{Sklyanin:1991ss,Sklyanin:1995bm}. Its key step is switching to variables in which the wavefunctions factorise and take the form of a product of building blocks known as $Q$-functions. These are fixed by solving a functional equation called the Baxter equation, which then often allows one to compute the spectrum. 

For our model we will show that, while the usual Bethe ansatz is rendered inapplicable in the conventional approach, the Baxter equation instead works perfectly, and truly reveals its power. Let us first briefly recall how this equation is formulated in the undeformed case. It reads
\beq
\label{bax1}
    (u+i/2)^JQ^{++}+(u-i/2)^JQ^{--}=\tau Q,
\eeq
where $Q$ is the $Q$-function while $\tau$ is the transfer matrix eigenvalue, and we used the shorthand notation
\beq
    f^\pm = f(u\pm i/2) \ , \qquad \ f^{[+a]}=f(u+ia/2).
\eeq
In this case, the $Q$-function $Q(u)$   is a polynomial built from Bethe roots $u_n$ as
\beq \label{eq:undef-Q}
    Q(u)=\prod_{n=1}^K(u-u_n) ,
\eeq
with $K$ the number of Bethe roots (magnons). 
In fact, imposing the Baxter equation and demanding that $Q(u)$ and $\tau(u)$ are polynomials uniquely fixes their form: one obtains a discrete set of solutions which correspond to the eigenstates of the spin chain. The associated energy   is then computed from
\beq
\label{evq}
E = \left.  \frac{i\lambda}{8\pi^2} \partial_u \log \frac{Q(u+i/2)}{Q(u-i/2)}\right\vert_{u=0}.
\eeq

Let us now discuss how this picture can be adapted to the Jordanian spin chain. Following the initial observations of \cite{Driezen:2025dww}, we propose that in fact both the Baxter equation \eq{bax1} as well as the expression \eq{evq} for the energy retain \textit{the same} functional form as in the undeformed case. 
The only modification is that  the Baxter equation now involves the eigenvalue of the  twisted transfer matrix $\tau^\star$, 
and the  deformation will enter the construction through its fixed leading coefficient (as we discuss in more detail below).
Although supporting evidence for this proposal was already presented  in \cite{Driezen:2025dww}, here we will be able to strengthen it considerably by showing that the Baxter equation perfectly reproduces the highly nontrivial direct calculations of the transfer matrix eigenvalues and energies done in section \ref{s:spin-chain}.  Furthermore, in section \ref{sec:strings} we will show that, at large $J$, the results from the Baxter approach   also reproduces both the classical ground state energy of the corresponding Landau-Lifshitz model \cite{Driezen:2025dww} and the semiclassical corrections derived from the algebraic curve analysis  of the corresponding string sigma-model \cite{Borsato:2022drc}.   Together, these finding leave little doubt about the validity of the proposal, and opens the way for us to efficiently compute the spectrum using the Baxter framework in a variety of settings.

In principle, in order to rigorously derive the Baxter equation one would need to fully implement the SoV approach together with an operatorial construction of the $Q$-operator. While this has been done for a variety of $sl(2)$ spin chain models (see e.g. \cite{Derkachov:2001yn,Derkachov:2002wz,Derkachov:2002tf,Bazhanov:2010ts}), it is a rather challenging task that we leave for the future. Nevertheless, there are also general reasons to expect that the resulting Baxter equation  remains the same as in the undeformed model. Indeed, for many different models based on rational $SL(2,R)$-type $R$-matrices the Baxter equation actually has a universal form, and the details of the model typically enter only through boundary conditions or analytic properties of the $Q$-functions. Concrete examples include diagonal twist deformations and even the rather drastic dipole deformation described in \cite{Guica:2017mtd}. Broadly speaking, this universality is also supported by the general idea that the Baxter equation should serve as a quantum version of the classical spectral curve of the integrable system, see e.g.~\cite{Bazhanov:1996dr, Chervov:2006xk}.\footnote{See \cite{Guica:2017mtd} for an example of how this logic helps to actually deduce the Baxter equation in a non-trivially deformed model.} That said, in our case the Yangian structure itself is also deformed in the sense that the usual $RTT$ relations which define the Yangian get replaced with $R^\F T^\star T^\star$ relations \eqref{eq:R*TFTF} (see also \cite{Kulish_1997}),\footnote{We point out that  the ``dipole equivalent'' \cite{Guica:2017mtd} of $R^\F T^\star T^\star$ reduces to undeformed $RT^\star T^\star$ relations (see also \cite{slavnov2019algebraicbetheansatz}). In contrast, in the Jordanian case no such simplifications appear  possible, and we verified that $R T^\star T^\star$ relations would be inconsistent with the $\alg{sl}(2)$ algebra in the Jordanian case, so that the Yangian structure is truly deformed.} so the fact that the Baxter equation still seems to remain unchanged is far from trivial.

In the next subsection we establish in more detail the key features of $Q$-functions for the Jordanian deformed case.

\subsection{Asymptotics and regularity of $Q$-functions}

\label{sec:qas}

As mentioned above, although the Baxter equation for the Jordanian model retains the same functional form as in the original spin chain, the deformation parameter $\xi$ will now enter through the modified leading coefficient of the transfer matrix eigenvalue. This seemingly innocent change in one coefficient in fact leads to drastic changes in the behaviour of the solutions and gives rise to a nontrivial set of new features.

\paragraph{Asymptotics and comparison.}  To appreciate the peculiar features arising due to the Jordanian deformation, it is instructive to compare it with three other models: the undeformed spin chain, the diagonally twisted spin chain, and the dipole deformation of \cite{Guica:2017mtd}. A summary of their key  characteristics is given in table \ref{tab:4ca}. In all cases, the transfer matrix is a polynomial of degree equal to the spin chain length $J$. While the functional form of the Baxter equation remains the same, the models differ in how the first two leading coefficients of the transfer matrix eigenvalue are fixed. The remaining coefficients are dynamical: they  depend nontrivially on the state, and can be determined from the Baxter equation by imposing appropriate regularity conditions on $Q$ (which select the physical solutions), as we will soon see for our Jordanian model as well.  Let us now briefly review the main features of each model.

\begin{table}[t]
\centering
\caption{Comparison of four spin chain models.}
\label{tab:4ca}
\begin{tabular}{|c|c|c|c|}
\hline
 {\bf Model} & {\bf Transfer matrix} & {\bf $Q$-functions}  \\
\hline
Undeformed & $\color{blue}2\;\color{black}\cdot u^J + \color{red}0\color{black}\cdot u^{J-1}+\dots$  & polynomial  \\
\hline
Diagonal twist & $\color{blue}2\cos\phi\;\color{black}\cdot u^J + \color{red}(?)\color{black}\cdot u^{J-1}+\dots$ & exponent times polynomial  \\
\hline
Dipole deformation & $\color{blue}2\;\color{black}\cdot u^J + \color{red}\beta\color{black}\cdot u^{J-1}+\dots$ & complicated non-polynomial  \\
\hline
Jordanian deformation & $\color{blue}2\cos\phi\;\color{black}\cdot u^J + \color{red}0\color{black}\cdot u^{J-1}+\dots$ & complicated non-polynomial \\
\hline
\end{tabular}
\end{table}

\begin{itemize}
    \item For the \textbf{undeformed} spin chain, the leading coefficient of 
the transfer matrix is fixed to 2, while the first subleading one vanishes. This ensures that there exist polynomial solutions for $Q$, and imposing polynomiality then fixes the remaining coefficients of the transfer matrix for each state. 

\item For the \textbf{diagonally twisted} spin chain, the leading coefficient is  deformed to $2\cos\phi$ where $\phi$ is the twist angle (see e.g. the review \cite{slavnov2019algebraicbetheansatz} and references therein).\footnote{The corresponding twist matrix is $\begin{pmatrix}
    e^{i\phi} & 0 \\ 0 & e^{-i\phi}
\end{pmatrix}$.} The physical solutions for $Q$ are then given by the exponent $e^{\phi u}$ times a polynomial. In this case, the first subleading coefficient of $\tau^\star$ at ${\cal O}(u^{J-1})$ is not fixed a priori but is instead determined dynamically, i.e.~its value will depend on the state. We indicate this by a question mark in the table \ref{tab:4ca}. However, importantly, this coefficient is always \textit{nonzero} -- since, as we will also see later in this section,  solutions for $Q$ given by exponents times polynomials would otherwise be excluded. 

\item After these two standard cases comes the \textbf{dipole deformation} studied in \cite{Guica:2017mtd}. Here the leading coefficient is undeformed (i.e.~set to 2), while the first subleading coefficient $\beta$ is \textit{nonzero} and  a  model-dependent parameter.\footnote{For the dipole case \cite{Guica:2017mtd}, $\beta=\xi M J$ with $\xi$ the deformation parameter and $M$ the  charge of the global root generator $\Delta^{J-1}(\e)$.}  This modification drastically affects  the $Q$-functions: they are no longer polynomials and their large-$u$ asymptotics is rather involved. For instance, for $J=2$ one finds $Q\sim u^{-3/4}e^{c \sqrt{u}}$ with $c$ a model-dependent constant.

\item Finally, we come to the \textbf{Jordanian deformation} studied in this paper. Here the situation is, in a sense, reversed compared to the dipole case:  the leading coefficient is now deformed, while the first subleading one remains undeformed (i.e.~set to zero) \cite{Driezen:2025dww}. The resulting $Q$-functions are again non-polynomial, and for instance at $J=2$ their large-$u$ behaviour  takes the form  $Q \sim u^{-1} e^{\pm \phi u}$ with $\phi=\tfrac{i}{2} \log(1+\xi M)$. 
\end{itemize}

To make the latter point more precise, let us consider the large $u$-asymptotics of the Jordanian case in general. 
Expanding the Baxter equation at large $u$ for arbitrary $J$ and making an ansatz for the asymptotic series of $Q(u)$, one has \cite{Driezen:2025dww}
\beq
    Q_a\sim  e^{\phi_a u}u^{-J/2}\left(1+\sum_{n=1}^\infty \frac{c_{n,a}}{u^n}\right) , \label{eq:Q_asymptotics}
\eeq
where the subscript $a=1,2$ indicates that the Baxter equation has two independent solutions with different asymptotics, with exponents $\phi_1=\phi, \ \phi_2=-\phi$. The true physical $Q$-function will be a particular linear combination of these two asymptotic series.
This behaviour immediately shows that the Jordanian $Q$-functions are neither polynomials nor simple exponentials times polynomials (as in the diagonally twisted case), but instead have a  more complicated  structure. Note however that  on the complex curve these asymptotics are much simpler than in the dipole case.

Finally, looking at table \ref{tab:4ca}, it is natural to ask whether a more general deformation might exist that renders \textit{both} leading coefficients of the transfer matrix arbitrary, i.e.~effectively combining the Jordanian and dipole deformations. While we are not immediately aware of a concrete spin chain realisation of such a  generalised model, the corresponding Baxter equation with these restrictions appears to be well defined, and we discuss it in section \ref{sec:gdef}.

\paragraph{Regularity.} 

The last crucial part of the Baxter equation approach is the regularity conditions to be imposed on the $Q$-functions in order to select the physical solutions. In the undeformed spin chain, this amounts to requiring polynomiality of $Q$, while in the diagonally twisted case one demands that $Q$ is an exponential times a polynomial. In both cases, imposing these analytic conditions fixes the transfer matrix eigenvalue and the $Q$-function itself, yielding a discrete set of solutions  labelled by quantum numbers  corresponding to the spin chain eigenstates. 
For the Jordanian model, however, the situation is more subtle:  the $Q$-functions can no longer be polynomials (even if multiplied by exponentials) due to their asymptotic behaviour \eqref{eq:Q_asymptotics}. 
It is therefore a nontrivial question 
what analytic requirement should replace polynomiality in order to
obtain a quantised, discrete spectrum of physical states.

The analysis of the following subsections will show that the correct requirement for the Jordanian model is analytic regularity of  $Q$, i.e.~that the physical solutions are those with no singularities for all finite values of $u$ (except possibly at infinity).\footnote{This can be seen as the minimal analytic requirement compatible both with polynomiality in the undeformed limit and with the deformed asymptotics,  motivated by  similar requirements in \cite{Guica:2017mtd}.} 
This is a crucial part which completes our proposal for how the Baxter equation works in the Jordanian model. 
In fact,  given the special large $u$ asymptotics, it is not even obvious that such solutions exist at all.
Remarkably, as we show below,  regular solutions do exist, and  imposing this condition   indeed restricts both $Q$ and the transfer matrix eigenvalue to a discrete set of solutions corresponding to the spin chain eigenstates. 
This regularity condition coincides with that used for the dipole deformation, where again the asymptotics  exclude any kind of simple function for $Q$ (however there the asymptotics are far more complex).

\subsection{Perturbative solution and spectrum}
Having discussed which form of the $TQ$-relation and corresponding regularity condition to demand, we proceed with the explicit computation of the spin chain spectrum  perturbatively in the deformation parameter. 

\subsubsection{$J=2$ spectrum} \label{ss:J=2-spectrum}
We begin with the $J=2$ Jordanian spectrum, which importantly also offers the simplest nontrivial setting to validate our Baxter approach through a direct comparison with the results of section \ref{s:spin-chain}.

As a warm up, we detail the computation of the deformed ground state energy ($S=0$).
From the generic asymptotics of the transfer matrix eigenvalue \eqref{eq:transfer_matrix_expansion}, one finds for $J=2$ 
\beq
    \tau^\star(u)=2\cos\phi \; u^2+ \tau^\star_0,
\eeq
with
\beq
    \phi=\frac{i}{2}\log(1+\xi M) .
\eeq
Rather than determining $\tau^\star_0$ via the ODE analysis of section \ref{ss:J=2-tau}, we now make a generic perturbative ansatz for $\tau^\star_0$  in the deformation object\footnote{Recall the remarks in footnote \ref{f:xi-and-M}.} $\xi M$ 
\beq
    \tau^\star_0=-\frac{1}{2}+\sum_{n=1}^\infty (\xi M)^n \tau^\star_{0,n} , 
\eeq
which reproduces, in the undeformed limit $\xi=0$,  the   solution 
for the $S=0$ state. 
The regularity condition on $Q(u)$ similarly allows for an expansion of the form
\beq
    Q_{S=0}(u)=1+\sum_{n=0}^\infty \sum_{k=1}^\infty q_{n,k}\; (\xi M)^nu^k.
\eeq
Here we fixed the overall normalisation of $Q(u)$ such that at ${\cal O}(u^0)$  the $Q$-function coincides with the one of the undeformed $J=2$ ground state, i.e.~$Q_{S=0,\xi=0}(u)=1 + {\cal O}(u)$. 
Substituting these ans{\"a}tze into the $TQ$-relation \eqref{bax1}  now uniquely fixes all coefficients $\tau^\star_{0,n}$ and $q_{n,k}$. Here, we show the results until ${\cal O}(\xi^7 M^7)$, with the deformed ground state $Q$-function given by\footnote{Note that $q_{n,k>n}=0$ always and  the undeformed $Q$-polynomial is recovered at $\xi=0$. }
\[\label{eq:q_function_ground_state}
  Q_{S=0}(u)={}& {1 - \xi^2 M^2  \frac{u^2}{24}  + 
  \xi^3 M^3 \frac{u^2}{24} - \xi^4 M^4 \left(  \frac{329 u^2}{8640} - \frac{ u^4}{1920}\right)}  +\xi^5  M^5 \left( \frac{149 u^2}{4320} - \frac{u^4}{960} \right) \\ 
  &{+ \xi^6  M^6 \left(- \frac{3905 u^2}{124416} + \frac{713 u^4}{483840} - \frac{u^6}{322560}\right) +{\cal O}(\xi^7 M^7),}
\]
while the resulting $\tau^\star (u)$ reproduces precisely the result \eqref{eq:J=2_deformed_eigenvalue} obtained by direct diagonalisation of the corresponding differential operator. Then, applying  the energy formula \eqref{evq}, we obtain
\[
\label{eq:energy_Baxter}
E_{S=0}=\frac{\lambda}{4\pi^2} \left(\frac{\xi^2 M^2}{24} - \frac{\xi^3 M^3}{24}+ \frac{131 \xi^4 M^4}{3456}-\frac{59\xi^5 M^5}{1728}+\frac{38441 \xi^6M^6}{1244160}+{\cal O}(\xi^7M^7)\right),
\]
which, as announced, matches exactly the ground state energy \eqref{eq:J=2_energy_check} obtained by direct diagonalisation of the Hamiltonian. 

In \cite{Driezen:2025dww} the same    ground state energy was obtained by inserting the transfer matrix eigenvalue derived from the ODE analysis into the Baxter $TQ$-equation. Here, by contrast, we solve the $TQ$-equation with generic regular ans\"atze for both $\tau^\star$ and $Q$, yielding the same energy in a self-contained and far more efficient manner. 

We now proceed analogously for the higher excited states by expanding in the deformation parameter around the undeformed Baxter pair $(\tau^\star_0, Q_S)$. For $J=2$,  this undeformed solution reads\footnote{See e.g.~\cite{Derkachov:2002tf} for this standard form in the $\mathfrak{sl}(2,R)$ spin chain. Here ${}_3F_2$ is the  generalised hypergeometric function, which for integer $S\geq 0$ truncates to a polynomial of degree $S$  in $u$, with $S$ Bethe roots.
}
\[ \label{{eq:tau-Q-undef=S}}
\tau^\star_0=-\frac{1}{2}-S(S+1) , \qquad Q_S(u) = {}_3F_2 (\{-S,S+1,iu + \frac{1}{2}\} , \{1,1 \} ,1) ,
\]
up to an overall normalisation. {Since $Q_S(u)$ is a polynomial of degree $S$ for integer $S\geq 0$, i.e.~it is always nonvanishing at ${\cal O}(u^S)$, we will  fix this freedom  by requiring that the coefficient of $u^S$ remains unity at all orders in $\xi$.\footnote{Note that this is very analogous to the normalisation condition used in the ODE analysis of section \ref{ss:J=2-tau}, cf.~around eq. \eqref{eq:normalisation_prescription}, and similarly reflects the freedom of multiplying the solution by an arbitrary $\xi$ dependant function.}}
For instance, for $S=1$ we take the ans{\"a}tze
\beq
    \tau^\star_0=-\frac{5}{2}+\sum_{n=1}^\infty (\xi M)^n \tau^\star_{0,n} , \qquad Q_{S=1}(u) = u+ {\sum_{n=0}^\infty \sum_{\substack{k\ge 0\\ k\neq 1}}^\infty q_{n,k}\; (\xi M)^nu^k} ,
\eeq
where the term $u^1$ is fixed to unity by normalisation.
Solving the $TQ$-relation  uniquely fixes all coefficients, yielding\footnote{Similarly as before,  one finds $q_{n,k>n+S}=0$ always  and the undeformed $Q$-polynomial is recovered at $\xi=0$. (up to normalisation). }
\[
&\tau^\star_0 =-\frac{5}{2}-\frac{7}{80}\xi^2 M^2 +\frac{7}{80}\xi^3 M^3+{\cal O}(\xi^4 M^4), \\ 
&Q_{S=1}(u) = u -\frac{\xi^2 M^2}{40}u^3 + \frac{\xi^3 M^3}{40} u^3 +{\cal O}(\xi^4 M^4) , \\
&E_{S=1}=\frac{\lambda}{4\pi^2}\left(2 h(1) + \frac{\xi^2 M^2 }{40}-\frac{\xi^3 M^3}{40}+{\cal O}(\xi^4 M^4)\right) .
\]
This reproduces precisely the results \eqref{eq:J=2_eigenvalue_all_S} with $S=1$ and \eqref{eq:J=2_energyy_S=1} from the direct diagonalisation of $\hat{\tau}^\star$ and the Hamiltonian. 
We have also confirmed \eqref{eq:J=2_eigenvalue_all_S} for $S=2,3$. Combined, these findings provide yet another strong confirmation   of our Baxter approach. 

Finally, for generic $S$, we generalise the ans{\"a}tze as discussed, i.e.
\[
\tau^\star_0=-\frac{1}{2}-S(S+1) +\sum_{n=1}^\infty (\xi M)^n \tau^\star_{0,n}, \qquad    Q_{S}(u) = u^S+ {\sum_{n=0}^\infty \sum_{\substack{k\ge 0\\ k\neq S}}^\infty q_{n,k}\; (\xi M)^nu^k}. 
\]
Substituting these into the $TQ$-relation and solving for   $S=1, \cdots, 20$ we can identify the $S$ dependence of   the resulting transfer matrix eigenvalue and obtain again perfect agreement with the closed-form result
 \eqref{eq:J=2_eigenvalue_all_S}. The corresponding   energy  reads 
\[ \label{eq:baxter-E-all-S}
E_S= \frac{\lambda }{4 \pi ^2}   2h(S) + \frac{\lambda}{32\pi^2} \left( \frac{S(S+1)-1}{4S (S+1)-3} \right)\xi^2 M^2 + {\cal O}(\xi^3 M^3).
\]
This expression constitutes the full  energy spectrum of the $J=2$ Jordanian-twisted spin chain. Notice that each undeformed level labelled by spin $S$ acquires a distinct deformation.
However, in the deformed theory, let us stress again that $S$ is only a convenient label for the resulting branches of states rather than a conserved quantum number, since for $\xi\neq 0$ the Hamiltonian no longer commutes with the two-site Casimir. 

\subsubsection{$J=2$ off-shell solution}

It is also instructive to consider the general perturbative solutions of the Baxter equation, without imposing regularity of the $Q$-functions. To solve the Baxter equation as a series in $\xi M$, we can use the versatile iterative method of \cite{Gromov:2015vua}. The solutions will be written in terms of $\eta$-functions, defined as 
\beq
    \eta_{s_1,\dots,s_k}(u)=\sum_{n_1>\dots>n_k\geq 0}^\infty\frac{1}{(u+in_1)^{s_1}\dots (u+in_k)^{s_k}}.
\eeq
We find the following two independent solutions to order $\xi^2M^2$
\[
    q_1&=1+\frac{1}{48}\xi^2M^2(-2u^2+i(48\tau_{0,2}^\star+1)\eta_1^+) \ ,  \\
   q_2&=\eta _{2}^++M^2 \xi ^2 \left(i\left( \tau_{0,2}^\star
+\frac{1}{48}\right) \left( \eta _{2,1}^+- \eta _{1,2}^+ \right)  +\frac{(-4 u^2-1)}{96}  \eta _{2}^+ +\frac{(1+2 i u)}{24} \right).
\]
Because of the $\eta$-functions, these solutions generally have poles as functions of $u$. In fact, $q_2$ is manifestly non-regular. Requiring that there exists a solution to order ${\cal O}(\xi^2 M^2)$ without any poles, here necessarily $q_1$,\footnote{In general, the regular solution can be a linear combination of $q_1$ and $q_2$. 
} we see that the coefficient of $\eta_1$ in $q_1$ must vanish. This condition would fix
\beq
    \tau_{0,2}^\star=-\frac{1}{48},
\eeq
exactly as we obtained earlier in \eq{eq:J=2_deformed_eigenvalue}. Consequently, we also see that  what remains of the regular $q_1$-function  is a polynomial in $u$ at each order in $\xi M$, just as  found before as well.

\subsubsection{Results for any $J$} \label{s:all-J-spinchain}
The $TQ$-relation is so powerful that it can also be employed  to derive the spectrum of the twisted spin chain at generic length $J$. We  present here the  explicit computation for  the ground state  and  $S=1$ excited states. These quantities are particularly important, as they allow to verify the string-spin conjecture by comparing to the spectra in  the large $J$ limit in section \ref{s:string-large-J}.

We consider the generic form for the eigenvalue of the transfer matrix  \eqref{eq:transfer_matrix_expansion}, which is valid for any $J$, and now make the following ansatz for the remaining coefficients
\beq\label{eq:tau_ansatz_all_J}
    \tau^\star(u)= 2 \cos \phi \ u^J+0\times u^{J-1}+\sum_{k=0}^{J-2} \sum_{n=0}^\infty u^k \phi^n \tau^{\star}_{n,k}.
\eeq
Here, we have traded the expansion parameter $\xi M$ for $\phi=\tfrac{i}{2} \log(1+\xi M)$, which turns out to be more convenient to studying the large $J$ regime.

In practice, we solve the $TQ$-relation as described above. To obtain the ground state energy we expand the $Q$-function around  $Q(u)=1+{\cal O}(u,\phi^0)$. Doing the calculation for $J=2, \cdots, 12$, we can identify the general $J$ dependence, giving
\[
Q_0(u)={}&1+\frac{u^2 \phi^2}{2(J+1)}+\frac{\phi^4}{8(J+1)(J+3)}\left(u^4 +\frac{Ju^2}{3(J+1)}\right)\\
    &+\frac{\phi^6}{48(J+1)(J+3)(J+5)}\left(u^6+\frac{J u^4}{J+1}+\frac{J(J(J+11)+9)u^2}{15(J+1)^2}\right)+O(\phi^8).
\]
From this result we   extract the ground state energy $E_0$ using  relation \eqref{evq}, giving
\[ \label{eq:E-J-S=0-O10}
  E_0=&-\frac{\lambda  }{8 \pi ^2 (J+1)} \phi ^2 -\frac{\lambda  }{96 \pi ^2 (J+1)^2}\phi ^4-\frac{ \lambda (J+3)  }{2880 \pi ^2 (J+1)^3} \phi^6\\
  &-\frac{\lambda\left(10 (J-1)^3+100 (J-1)^2+345 (J-1)+408\right)   }{483840 \pi ^2 (J+1)^4 (J+3)} \phi ^8 +{\cal O}(\phi^{10}).
\]
Notably, this result is even in $\phi$. 
At large $J$ and fixed $\xi$, $M$ and $\lambda$,  it  behaves as\footnote{This expansion shows how the use of  $\phi$ instead of $\xi M$ is particularly convenient in this regime: it resums an infinite series of higher-order $\xi M$ contributions into a compact form, so that the leading ${\cal O}(J^{-1})$ term is exact in $\phi$. This structure would be obscured if one expanded directly in $\xi M$.}
\[
E_{0}=-\frac{\lambda }{8\pi ^2J }\phi^2 - \frac{\lambda \phi^2 (-12 + \phi^2 ) + {\cal O}(\phi^6)}{96\pi^2 J^2} +{\cal O}(J^{-3}) , \label{eq:E0-J-LL-match}
\] 
where the subleading terms $ {\cal O}(J^{-2})$ receive contributions from all even powers $\phi^{2n}$ with $n \geq 1$. Interestingly, the leading ${\cal O}(J^{-1})$ term is exact in $\phi$ and  reproduces precisely
 the classical ground state energy derived in the continuum (Landau-Lifshitz) regime of the spin chain with twisted boundary condition, cf.~eq.~(4.26) of \cite{Driezen:2025dww}.
 This confirmation thus provides  a nontrivial consistency check between the use of the $TQ$-relations and  the explicit twisted-boundary conditions of \cite{Driezen:2025dww}.   

To compute the  energy of the first excited states  for generic $J$,  we can keep the ansatz \eqref{eq:tau_ansatz_all_J} for $\tau^\star(u)$ and expand the $Q$-function  around the undeformed $S=1$ one-magnon solution, i.e. 
\[
Q_{1,n}(u)=u-u_n + \sum_{k=1}^\infty \phi^{k} c_{k} +  \sum_{k=0}^\infty \phi^{k} \sum_{l=2}^\infty u^l c_{k,l}  ,
\]
where  $u_n=\frac{1}{2} \cot\frac{\pi n}{J}$ is the undeformed Bethe root  associated with the $n$-th magnon mode of momentum $p_n = \frac{2\pi n}{J}$, with $n=0,1\cdots,J-1$. We however exclude the $n=0$ mode, as it corresponds to a root at infinity with vanishing energy, which decouples from the other excitations.\footnote{Note that for $J=2$ the only independent physical mode is $n=1$ with $u_1=0$.}  Solving the $TQ$ system for $J=2,\cdots ,  19$ and identifying the $J$ dependence, we find the $Q$-function until ${\cal O}(\phi^2)$
\[
Q_{1,n}=u-u_n+ \frac{u  \left((J+2) u^2-(J+6) u u_n-u_n\right)}{2 (J+2)(J+3)}\phi^2+{\cal O}(\phi^4) ,
\]
and the corresponding
 energy  until ${\cal O}(\phi^4)$
\[ \label{eq:sc-E1-allJ}
E_{1,n}={}& \frac{\lambda}{2\pi^2 (1+4 u_n^2)} - \lambda\frac{ J+2 + 4 (J+6) u_n^2 }{8\pi^2(1+4u_n^2)(J+3)(J+2)} \phi^2 \\
&- \lambda\frac{ (J+2)^3 (J^2 + 8J + 9)  + 4 (J+2)^2 (J (J (J+14) + 135) +270) u_n^2  }{96\pi^2 (1+4 u_n^2) (J+5)(J+3)^3(J+2)^3 }\phi^4 \\
&- \lambda\frac{   (96 J (J(J+20) + 48) ) u_n^4 }{96\pi^2 (1+4 u_n^2) (J+5)(J+3)^3(J+2)^3 }\phi^4 +{\cal O}(\phi^6).
\]
These expressions reproduce the $J=2$ results of section \ref{ss:J=2-spectrum}.
In  section \ref{sec:strings} we will match the continuum limit  ($J\rightarrow\infty$) of the above energies   to leading and subleading order in $1/J$ with the semiclassical one-loop spectrum of  the corresponding string solution (under a particular identification of the  residual momentum charges on both sides) given a highly nontrivial check of  the Baxter framework as well as Jordanian-deformed AdS/CFT.

\subsection{Numerical solution for $J=2$}

The results of the previous sections are all perturbative expansions in the deformation parameter $\xi$ (or, equivalently, $\phi$) that probe only the small deformation regime. The power of the Baxter framework can however also be employed to investigate the intermediate and large deformation regime.
In particular, here we will show how to obtain the spectrum of the $J=2$ Jordanian chain numerically by solving the Mellin transform of the Baxter equation.\footnote{Our regular $Q$-function has exponential asymptotics with an imaginary twist, which means they grow fast in both the upper and lower half plane. Therefore simpler methods  of numerically solving the Baxter equation, like \cite{Gromov:2015wca}, seem to be not applicable here. }

\paragraph{Mellin transform of the $TQ$-relation.} The Mellin transform allows one to convert finite difference equations into differential ones, making it a very suitable tool to tackle the $TQ$-relation (see e.g. an early discussion in \cite{Faddeev:1996iy}). We follow the conventions of appendix D of \cite{Beccaria:2009rw} for the Mellin transform, which reads
\[
Q(u) = K \int_0^{\infty} d\omega \, {\omega}^{iu-1} \hat{Q}(-\omega)\,, \quad
K = \frac{1}{\Gamma(iu)\Gamma(1-iu)}. \label{eq:Mellin_transform}
\]
It satisfies the following properties, 
\[
(u+bi)^L \, Q(u)
=
K \int_0^{\infty} d\omega \,
{\omega}^{iu-1} \, {\omega}^{-b} \,
{\left[  i\omega \frac{d}{d\omega} \right]}^L
\left\{ {\omega}^b \hat{Q}(-\omega)\right\}
, 
\label{eq:Mellin_crucial_property1}
\]
and
\[
(u+bi)^L \, Q(u\pm i)
=  K \int_0^{\infty} d\omega \,
{\omega}^{iu-1} \, {\omega}^{-b} \,
{\left[  i\omega \frac{d}{d\omega} \right]}^L
\left\{ -{\omega}^{b\mp 1} \hat{Q}(-\omega)\right\}
, 
\label{eq:Mellin_crucial_property}
\]
which allow us to convert the $TQ$-relation into a differential equation. Taking the $J=2$ $TQ$-relation \eqref{bax1} and applying the Mellin transform, together with $\omega \rightarrow-\omega$, then gives
\[ \label{eq:mellin_transformed_TQ_relation}
4 \omega  \Big(\left(-2\cos(\phi) \omega +\omega ^2+1\right)& \hat{Q}''(\omega)+\left(2 \omega -2\cos(\phi)\right) \hat{Q}'(\omega)\Big)\\
&+\hat{Q}(\omega) \left(4 \tau _0^\star+\omega +\frac{1}{\omega }\right)=0.
\]
This ODE has two regular singular points at $\omega=0$ and $\omega\rightarrow\infty$. To proceed further, we perform the  change of variables $\omega =-\frac{z}{1-z}$ together with the redefinition $\hat{Q}(z)=\sqrt{z(1-z)}\Psi(z)$,
such that in total 
\[
Q(u)=\frac{i\sinh{\pi u}}{\pi}\int_0^1dzz^{iu-1/2}(1-z)^{-iu-1/2}\Psi(z).\label{eq:total_mellin}
\]
After these redefinitions, eq.~\eqref{eq:mellin_transformed_TQ_relation} gets modified straightforwardly but to a lengthy ODE which we do not report here for brevity. Importantly, the changes of variables maps the singular points to $z=0,1$, which makes the ODE more suited to numerical solving.

\begin{figure}[t!]
\begin{center}
 \centerline{\includegraphics[width=14cm]{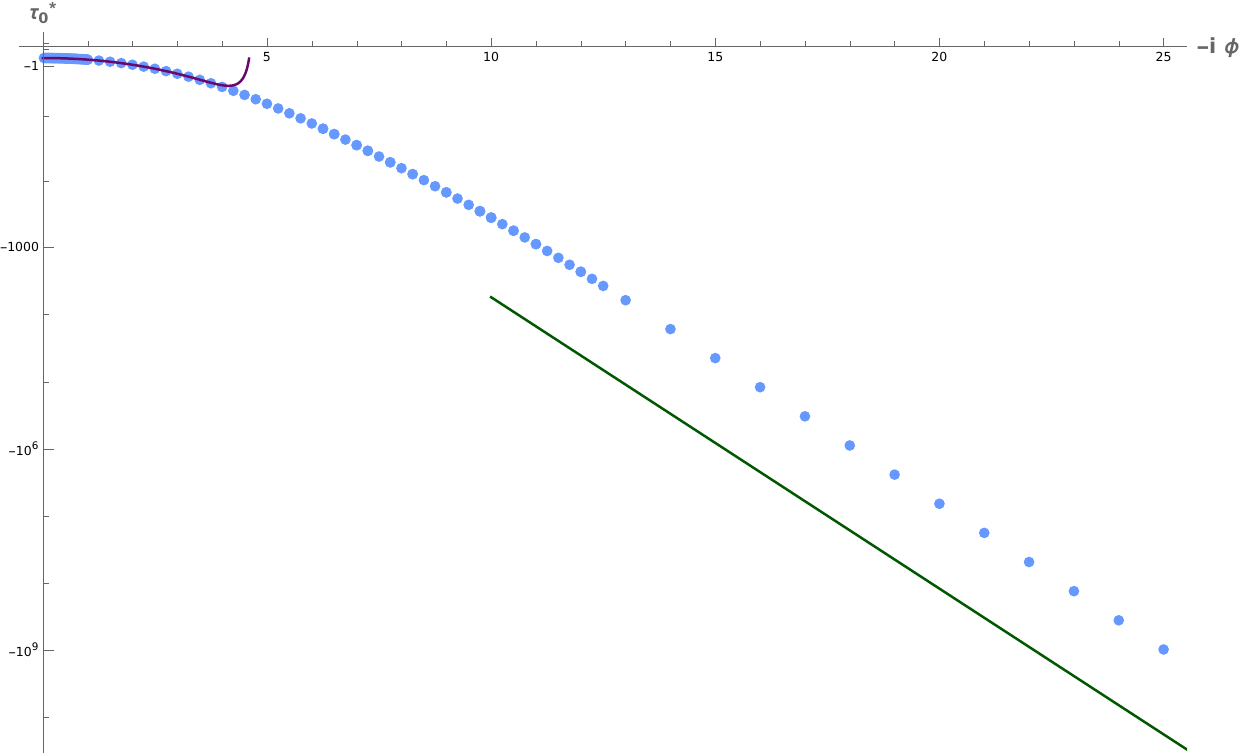}}
\caption{\label{fig:t0_numerics}\small The ground state ($S=0$) transfer matrix eigenvalue as a function of the deformation parameter $-i\phi$. The blue points are the numerical results. The purple curve at small $\phi$  is the 20-th order small-$\phi$ expansion obtained analytically from solving the $TQ$-relation. The green curve is a large $\phi$ asymptotic approximation, $\tau_0^\star=-\frac{1}{2} \cos(\phi)$, obtained from a preliminary WKB analysis, the details of which will be discussed elsewhere.}
\end{center}
\end{figure}

\paragraph{Numerics.} We can now discuss numerical results for the $J=2$ spectrum for intermediate and large deformation regime of the Jordanian chain. The Mellin transformed $TQ$-relation \eqref{eq:mellin_transformed_TQ_relation} (after change of variables) has two singular points at $z=0,1$, which lie at the boundary of the integration in \eqref{eq:total_mellin}. At a given value of $\phi$, we solve this ODE while imposing regularity of the solution at these two singular points\footnote{Heuristically, $\hat{Q}$ has logarithmic singularities at its singular points, which would be converted by \eqref{eq:total_mellin} into singularities of $Q$ at $u= \mathbb{Z}i/2$  (see also \cite{Guica:2017mtd}), which we require vanish. }: very explicitly, at $\phi$ fixed, we sample over values of $\tau_0^\star$ and run a numerical solving of the ODE over $z\in [\varepsilon,1-\varepsilon]$ for each value of $\tau_0^\star$ in the sampling, where $\varepsilon$ acts as a regulator for the singular points. We then numerically impose regularity at the boundaries,\footnote{In practice, we minimize $(\varepsilon \Psi'(\varepsilon))^2 + ((1-\varepsilon)\Psi'(1-\varepsilon))^2 $, which would guarantee regularity if it vanished.} which singles out a value for $\tau_0^\star$. We iterate this procedure over some range in the deformation parameter $\phi$, which produces figure \ref{fig:t0_numerics}. 

\begin{figure}[t!]
\begin{center}
 \centerline{\includegraphics[width=14cm]{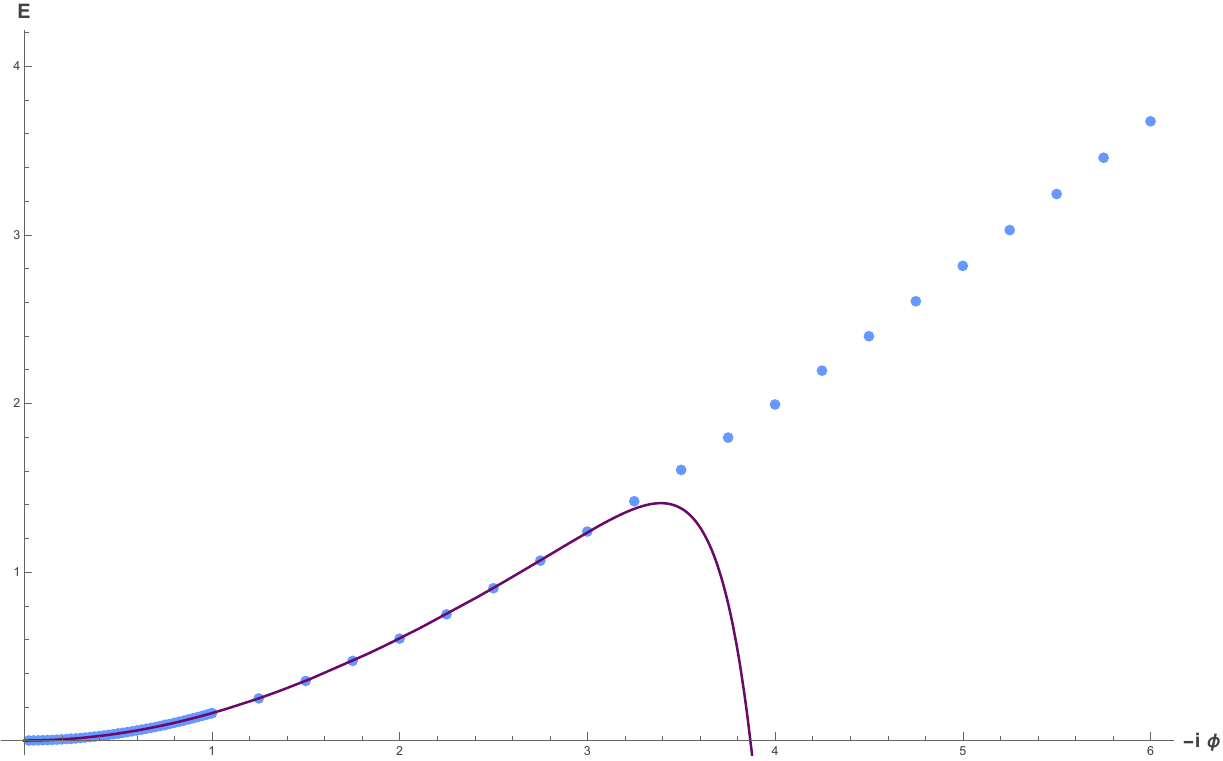}}
\caption{\label{fig:E_numerics}\small The ground state ($S=0$), $J=2$ energy as a function of the deformation parameter $-i\phi$. The blue points are the numerical results. The purple curve is the 20-th order small-$\phi$ expansion obtained analytically from solving the $TQ$-relation and evaluating the energy formula \eqref{evq}.}
\end{center}
\end{figure}

Then, given the numerical solution of \eqref{eq:mellin_transformed_TQ_relation} at fixed $\phi$ and with corresponding value of $\tau_0^\star$, we can perform the inverse Mellin transform through a numerical integration to obtain a numerical solution for the Baxter $Q$-function at a given value of $\phi$ . Again we regulate the singular points at the integration boundaries with $\varepsilon$. We can then insert this solution into the energy formula \eqref{evq} to obtain the energy, and threading this procedure over several values of $\phi$, we obtain figure \ref{fig:E_numerics}.

\paragraph{Discussion.} In both figure \ref{fig:t0_numerics} and \ref{fig:E_numerics}, we find very good agreement between the numerical solving and the analytic perturbative solution of the $TQ$-relation, which for this analysis was computed up to ${\cal O}(\phi^{20})$. These analytical perturbative expansions stay valid only within the small deformation range (up to $\phi\approx3.5 i$), but our numerical results smoothly extend past this point, demonstrating the  power of the Baxter framework across the full deformation range.

In figure \ref{fig:t0_numerics}, we also reported an analytic result for $\tau_0^\star$ in the large $\phi$ regime (green curve), which resulted from a preliminary WKB analysis of \eqref{eq:mellin_transformed_TQ_relation}. The numerical trend matches this analytical behaviour, and in fact, numerical analysis up to $\phi= 400 i$  confirms this prediction. We hope to discuss this WKB analysis in more detail elsewhere.

As discussed, the numerical integration associated to the inverse Mellin transform \eqref{eq:total_mellin} requires regulating the integration boundaries. This turns out to be a numerical bottleneck for the range of validity in $\phi$ of the energy computation, as more and more contributions to the integral are missed closed to the singular points.\footnote{See also \cite{Guica:2017mtd}, section 6.3 for a detailed discussion of a similar situation, with analytical results. There, at large deformation, the integration was dominated by its endpoints and we expect a similar behaviour to be relevant here.} This explains the restricted range in figure \ref{fig:E_numerics} relative to figure \ref{fig:t0_numerics}. However, this bottleneck does not affect the numerical computation of $\tau^\star_0$, and in fact, we have computed $\tau_0^\star$ up to $\phi=400 i$ without observing change in its behaviour, so that we do not expect a change in behaviour for the energy at larger deformation.

\subsection{The most general 2-parameter deformation}
\label{sec:gdef}

In section \ref{sec:qas} we compared different deformations that correspond to having different nontrivial coefficients in the transfer matrix. Here we briefly discuss one case not included there; namely, when the leading and subleading coefficients are \textit{both} fixed to generic values. While we do not consider the corresponding deformed spin chain here, we will explore what happens to the $Q$-functions under the analytic regularity condition. 

For this ``most general'' deformation the transfer matrix eigenvalue $\tau^\star (u)$ reads
\beq
\label{tgen}
    \tau^\star(u)=(2+\alpha)u^J+\beta u^{J-1}+\sum_{k=0}^{J-2}\tau^\star_ku^k ,
\eeq
where $\alpha,\beta$ are arbitrary finite parameters, with $\alpha=\beta=0$ giving the undeformed case. The Jordanian deformation, the dipole deformation, and the diagonal twist deformation\footnote{In the diagonal twist case $\beta$ is not actually a free parameter, rather it is fixed by imposing the additional regularity requirement that $Q$ is an exponent times polynomial.} are all special cases of this expression \eq{tgen}. Let us show explicitly for $J=2$ how we can solve the Baxter equation as an expansion in $\alpha,\beta$ that we assume to be perturbative small parameters (with their ratio $\alpha/\beta$ kept fixed). We write 
\beq
\tau^\star_0=-\frac{1}{2}+\sum_{n=1}^\infty\tau_{n,0} ,
\eeq
with $\tau^\star_{n,0}\sim \alpha^n$, and we take, for the deformed vacuum state,
\beq
    Q=1+{\cal O}(\alpha) .
\eeq
Remarkably, demanding that $Q$ is polynomial at each perturbative order again fixes both $Q$ and the transfer matrix uniquely. We get e.g.
\beq
    \tau^\star _{1,0}= -\frac{\alpha }{12},\ \tau^\star _{2,0}= \frac{\alpha ^2}{270}+\frac{\beta ^2}{24},\ \tau^\star
   _{3,0}= -\frac{4 \alpha ^3}{8505}-\frac{\alpha  \beta ^2}{108},\ \tau^\star _{4,0}=\frac{1157 \alpha ^4}{15309000}+\frac{7 \alpha ^2 \beta
   ^2}{3240}+\frac{\beta ^4}{17280} ,
\eeq
and
\beq
    Q=1-\frac{1}{6}  u (3 \beta +\alpha  u)+\frac{ u \left(165 \alpha  \beta +18 \alpha ^2 u^3+120 \alpha  \beta  u^2+34 \alpha ^2 u+180 \beta ^2
   u\right)}{2160}+\dots
\eeq
The fact that a regular solution for $Q$ exists for generic values of $\alpha,\beta$ (and that imposing its regularity fixes the remaining transfer matrix coefficients) is in principle quite nontrivial. It can be viewed as a consistency check for this more general deformation, indicating that it indeed may be possible to realise it at the operatorial spin chain level. In the future it would be very interesting to understand how this can be done.

\section{Matching with the semiclassical string spectrum}
\label{sec:strings}

The Baxter analysis has provided the complete deformation of the ground and first excited spin chain states, valid for arbitrary length $J$.
We now turn to the string side of the correspondence to verify how these results emerge from the semiclassical spectrum of the Jordanian-deformed $AdS_5\times S^5$ string. In particular, we summarise the key features of the Jordanian worldsheet spectrum in the $\mathfrak{sl}(2,R)$ sector and  demonstrate how it reproduces the spin chain results in the large $J$ limit, thereby establishing the bridge between the discrete spin chain picture and the continuous worldsheet counterpart.

\subsection{Relevant ingredients of the Jordanian string worldsheet}

On the worldsheet,  Jordanian deformations are a subclass of Homogeneous Yang-Baxter (HYB) deformations \cite{Klimcik:2008eq,Delduc:2013fga,Delduc:2013qra,Kawaguchi:2014qwa,vanTongeren:2015soa} which are known to be on-shell equivalent to the undeformed string sigma-model with twisted boundary conditions \cite{Borsato:2021fuy}. 
This is  precisely the worldsheet counterpart of the boundary twist that reorganised the Jordanian $\mathrm{XXX}_{-1/2}$ chain. See \cite{Driezen:2025dww} for a review on these aspects with  the conventions and definitions used in this paper.

Our setting is the Green-Schwarz string on the supercoset
$PSU(2,2|4)/(SO(1,4)\times SO(5))$ deformed as a HYB deformation \cite{Klimcik:2008eq,Delduc:2013fga,Delduc:2013qra,Kawaguchi:2014qwa,vanTongeren:2015soa} by a rank-2 Jordanian $r$-matrix generated by
\[ \label{eq:def-jordanian-R}
\h=\frac{D-J_{03}}{2} , \qquad \e=p_0+p_3 \ , 
\]
with $D$, $J_{ij}$, and $p_i$ the dilatation, rotation and translation generators of the conformal $SO(2,4)$ subgroup of ${PSU}(2,2|4)$.
This is the unique  case with support confined to an $\mathfrak{sl}(2, R)$ subalgebra through a single parameter and with a constant dilaton configuration $\Phi(x)=\Phi_0$   \cite{Driezen:2025dww}.\footnote{The above $r$-matrix  is however non-unimodular, meaning that the corresponding background solves only the    ``modified'' supergravity equations \cite{Borsato:2016ose}. The model does have a   unimodular completion, obtained by extending the $r$-matrix to include supercharges \cite{2004Tolstoy,vanTongeren:2019dlq,Borsato:2022ubq}, thereby producing a genuine type IIB supergravity solution (see also \cite{Kawaguchi:2014fca,Matsumoto:2014ubv,Driezen:2024mcn}).
Nevertheless, since these additional  contributions only affect worldsheet interactions involving fermions, and the semi-classical spectral curve depends solely on the classical (bosonic) interactions (dictated by the spacetime metric and NSNS field)
there is no need for us to burden the discussion with such additional details, and we will restrict attention to the purely bosonic truncation.} 

Because  the $r$-matrix involves only $\mathfrak{so}(2,4)$ generators, the deformation acts nontrivially on the $AdS_5$ factor of the spacetime, while it leaves the $S^5$  untouched.  In  global $AdS_5$ coordinates
$(T,V,P,\Theta , Z)$ with ranges $T,V\in (-\infty, +\infty)$, $Z,P \in (0, +\infty)$ and $\Theta \in [0,2\pi)$ the deformed   metric and $B$-field read
\[ \label{eq:jor-G-B}
ds^2 &= \frac{dZ^2 + dP^2 + P^2 d\Theta^2 - 2 dT dV}{Z^2} - \frac{(4 Z^4+\eta^2) (Z^2+P^2)}{4 Z^6}dT^2 + ds_{S^5}^2 , \\
B &= \eta \frac{P dP \wedge dT}{2 Z^4}  - \eta d\left(\frac{dT}{4 Z^2}\right) ,
\]
where $ds_{S^5}^2$ is the metric of the  round $S^5$ and $\eta \in \mathbb{R}$  is the deformation parameter. 
It was shown in \cite{Borsato:2022drc} that these coordinates are geodesically complete and that the resulting geometry has Schrödinger-like scaling symmetries (see also \cite{Blau:2009gd}).  
They are furthermore  particularly convenient as they are adapted to the  residual commuting $\mathfrak{so}(2,4)$ symmetries of the deformed sigma-model,  
given by
\[
\mathfrak t_{\mathfrak{a}}=\mathrm{span} \{\, D + J_{03},\ k_0 + k_3,\ p_0 - p_3,\  J_{12} ,\ p_0 + p_3\, \}\ \cong\ \mathfrak{sl}(2,  R) \oplus \mathfrak u(1)^2.
\]
Of these, the   maximally commuting subalgebra relevant for the string spectral problem is generated by \cite{Borsato:2022drc}
\begin{equation} \label{eq:res-commuting-alg}
\begin{alignedat}{4}
&&H_T=\frac12(p_0 - k_0 - p_3 - k_3),\quad 
&&H_\Theta=J_{12},\quad 
&&H_V=\e= p_0 + p_3,
\end{alignedat}
\end{equation}
where $H_T$ is a  time-like generator for the above coordinate $T$ (with conjugate energy $\E$),
 $H_\Theta$ generates rotations in the angle $\Theta$ (with  spin $\S$)  and $H_V = \e$ is a \textit{null} isometry\footnote{In a compact representation $H_V$ is non-diagonalisable.} along the coordinate $V$,  whose conjugate charge has the interpretation of  {null momentum $\M$}.\footnote{Both $H_\Theta$ and  $H_V$ also commute with $D+J_{03}$, which is a radial-like scaling generator. Its conjugate has the interpretation of a non-relativistic scaling dimension and its eigenbases is related to that of $H_T$ via a non-relativistic state-operator map. See e.g.~\cite{Son:2008ye,Balasubramanian:2008dm} for more details and also  \cite{Guica:2017mtd} for  the Schr\"odinger/dipole deformation which has the same spectral interpretation  of the residual maximally commuting isometry algebra  of  AdS.}

This structure should be contrasted with the undeformed $AdS_5 \times S^5$ case, where the  isometries are large enough to represent the maximally commuting subalgebra by a (diagonalisable) Cartan subalgebra. It is generated by
two rotations,  for instance ($J_{12}$, $J_{34}$) with spins $(S_1, S_2)$, and either the time-translation generator $p_0 - k_0$ (energy $\Delta$) or the dilatation generator $D$ (relativistic scaling dimension), whose eigenbases are related by the standard relativistic state–operator map.

\subsection{Asymptotics of the twisted algebraic curve}

HYB deformations are  Lax-integrable \cite{Klimcik:2008eq,Delduc:2013fga,Delduc:2013qra,Delduc:2014kha}, which in principle enables a study of its semiclassical spectrum through the algebraic curve associated with the monodromy matrix of the one-parameter family of flat Lax connections. In practice, however, such an analysis requires identifying a suitable asymptotic regime of the curve from which the string energy $\E$ can be extracted.
This identification is only possible in the undeformed yet twisted formulation of the sigma-model \cite{Borsato:2021fuy,Borsato:2022drc},  highlighting again that, just as in the spin chain realisation, 
 the twisted-boundary formulation is  crucial for making the spectral structure of these deformations explicit.

For comparison, consider first the algebraic spectral curve for strings on undeformed $AdS_5\times S^5$, which is characterised by eight quasimomenta 
$\{\,\hat p_i,\tilde p_i \ ; \ i=1,\cdots, 4 \,\}$ living on an eight-sheeted Riemann surface parametrised by the complex spectral parameter $u\in \mathbb{C}$. The $\hat p_i$ describe the AdS sector, while the $\tilde p_i$ correspond to the sphere. 
Their asymptotic large-$u$ behaviour encodes the global charges associated to the commuting isometries.    In particular, after conjugation to the Cartan subalgebra generated by $\{p_0-k_0,J_{12},J_{34}\}$ in the AdS sector and  the three commuting rotations in the sphere sector, one finds   (see e.g.~\cite{Beisert:2005bm,Gromov:2007aq})
\[ \label{eq:large-u-undef-p}
\begin{pmatrix}
 \hat p_1 \\ \hat p_2 \\ \hat p_3 \\ \hat p_4 
\end{pmatrix} \sim 
\frac{2\pi}{u \sqrt{\lambda}}
\begin{pmatrix}
-\Delta-S_1 - S_2 \\ -\Delta+S_1+S_2 \\ \Delta+S_1-S_2 \\  \Delta-S_1 + S_2
\end{pmatrix}  , \qquad
\begin{pmatrix}
 \tilde p_1 \\ \tilde p_2 \\ \tilde p_3 \\ \tilde p_4 
\end{pmatrix} \sim 
\frac{2\pi}{u \sqrt{\lambda}}
\begin{pmatrix}
J_1 +J_2-J_3\\ J_1-J_2+J_3 \\ -J_1 + J_2+J_3 \\  -J_1-J_2-J_3
\end{pmatrix}  + {\cal O}(u^{-2}) ,
\]
where $\lambda$ is the 't Hooft coupling and $J_{1},J_2,J_3$ are the  angular momenta on $S^5$. 

After the Jordanian twist of the type \eqref{eq:def-jordanian-R}, the asymptotic behaviour of the  AdS  quasimomenta is drastically modified \cite{Borsato:2022drc}
\[ \label{eq:large-u-Jor-p}
\begin{pmatrix}
 \hat p_1 \\ \hat p_2 \\ \hat p_3 \\ \hat p_4 
\end{pmatrix} \sim \frac{i}{2 \sqrt{\lambda}} \begin{pmatrix}
0 \\ - \mathbf{Q} \\ \mathbf{Q} \\ 0
\end{pmatrix} + 
\frac{2\pi}{u \sqrt{\lambda}}
\begin{pmatrix}
-\E - \S\\ \S \\ \S \\  \E-\S
\end{pmatrix} + {\cal O} ( u^{-2}) ,
\]
while the $\tilde p_i$ remain as in the undeformed case. Here 
$\mathbf{Q}$ is a  ``twist-charge'' defining the twisted boundary conditions of the string.\footnote{Comparing with \cite{Borsato:2022drc,Driezen:2025dww} and other works note that we have rescaled  $\mathbf{Q}$ with a factor of $\sqrt{\lambda}$ to treat it on the same (classical) footing as the other charges $\E,\M,\S$.  }  Although $\mathbf{Q}$ is not direcly linked  to a generator of \eqref{eq:res-commuting-alg} in a usual (Noether) way,   it is conserved \cite{Borsato:2021fuy} and can be understood as being associated to the null isometry generated by $H_V=\e$.\footnote{Explicit  expressions clarifying this relation can be found in \cite{Borsato:2022drc,Driezen:2025dww} (see also \cite{Driezen:2024mcn}).} In this sense, $\mathbf{Q}$ provides the spectral-curve counterpart of the null momentum charge $\M$. 
Crucially, the Jordanian twist thus acts nontrivially on the $\hat2$-$\hat3$ pair by introducing a constant $u^0$ term in their asymptotics, while $\hat p_1,\hat p_4$ retain the usual $u^{-1}$ behaviour.
For the semiclassical spectra on solutions with $\S=0$, the structure of \eqref{eq:large-u-Jor-p} further indicates that the  $\mathfrak{sl}(2,R)$ polarisations on the string side---expected to map to one-magnon excitations of the  twisted $\mathrm{XXX}_{-1/2}$ spin chain---are  realised by  microscopic cuts connecting the  $\hat2$ and $\hat3$ sheets, as we will now discuss.

\subsection{Curve of the BMN-like solution}
The classical vacuum solution of the sigma-model equations of motion, which should correspond to the ground state of the twisted $\mathrm{XXX}_{-1/2}$ spin chain, is the ``BMN-like'' point-like string with vanishing spin $\S=0$ and fields propagating as
\[ \label{eq:BMN-sol}
T = - \kappa \tau , \qquad V = \frac{\eta^2}{4} m \tau  , \qquad Z = \sqrt{\frac{\kappa}{m}}  , \qquad P=0 , \qquad \psi = \cJ \tau , 
\]
where $\tau$ is the worldsheet time coordinate and $\psi$ an angle that parametrises a big circle in $S^5$.  The real parameters $\kappa, m , \cJ$ of the solution determine the non-zero classical charges 
\[ \label{eq:bmn-like-charges}
\E = \sqrt{\lambda} \kappa , \qquad \M=\sqrt{\lambda} m , \qquad J = \sqrt{\lambda} \cJ ,
\]
with $J$ the angular momenta for $\psi$ that corresponds to the spin chain length \cite{Beisert:2010jr}, following the notation used in the previous sections. The parameters 
 are related through the Virasoro constraint as
\[  \label{eq:on-shell-class-energy}
\kappa^2 = \cJ^2 + \frac{\eta^2 m^2}{4} ,
\]
which requires $\lvert {\eta m} \rvert \leq 2 \kappa$ and $\lvert \cJ \rvert \leq \kappa$ to ensure that all charges remain real. Classically, this implies the  off-shell  charge relation
\[ \label{eq:class-energy}
\E = \sqrt{J^2 + \frac{\eta^2 \M^2}{4}} .
\]
For $\eta=0$ this is the BMN/BPS vacuum relation  $\Delta = J$.

Because the BMN-like solution \eqref{eq:BMN-sol} is point-like, the  Lax connection is independent of the worldsheet space coordinate and the monodromy matrix can be explicitly diagonalised. Its resulting AdS and sphere quasimomenta on the full curve  are \cite{Borsato:2022drc}
\begin{equation} \label{eq:BMN-like-curve}
\begin{aligned}
\hat p_1 (u) &= - \hat p_4 (u) =   \frac{ 2\pi\kappa \sqrt{ u^2 - \tfrac{\eta^2 m^2}{4\kappa^2} }}{1-u^2} , \\
 \hat p_2 (u) &= - \hat p_3 (u) =  \frac{2\pi \kappa u \sqrt{1 - \tfrac{\eta^2 m^2}{4\kappa^2}  u^2}}{1-u^2} , \\
\tilde p_1 (u) &= \tilde p_2 (u) = - \tilde p_3 (u) = -\tilde p_4 (u) =  \frac{2\pi\cJ u}{u^2-1} .
\end{aligned}
\end{equation}
which we evaluated on the ``physical branch'' $\lvert u \rvert > 1$.\footnote{The complementary domain $\lvert u \rvert \leq 1$ is related to  $\lvert u \rvert >1$ by the inversion symmetry $u\mapsto u^{-1}$, which  follows from the underlying  $\mathfrak{psu}(2,2|4)$ structure of the sigma-model and simply relabels the quasimomenta \cite{Beisert:2005bm,Borsato:2022drc}.} Matching with \eqref{eq:large-u-Jor-p} in the asymptotic $u\rightarrow \infty$ regime confirms \eqref{eq:bmn-like-charges}   with $\S=0$ and gives the classical relation
\[ \label{eq:Q-to-M}
\mathbf{Q}= 2\pi \eta \sqrt{\lambda} m.
\]
which is consistent with the reformulation of the  BMN-like point-like solution in the twisted model, see \cite{Borsato:2022drc}.

The square-root structure of the AdS quasimomenta determines the branch cuts of the BMN-like spectral curve. Specifically, the branch points in $\hat p_{1,4}$ and $\hat p_{2,3}$ are fixed at $u=\pm \tfrac{\eta m}{2\kappa}$ and $u=\pm \tfrac{2\kappa}{\eta m}$ respectively. With $\lvert \eta m \rvert \le 2\kappa$, the former lie inside $\lvert u\rvert=1$ and the latter outside. We then adopt the standard algebraic curve convention to connect each real pair of branch points by a real cut along which the argument of the square root is negative.\footnote{
This ensures that the quasimomenta become purely imaginary across the cut, 
such that their associated densities are real. 
Note that this choice 
also  avoids crossing the simple poles at $u=\pm1$.
} The resulting configuration consists of two cuts  ${\cal C}_{\hat 1, \hat4} = [-\tfrac{\eta m}{2\kappa} , \tfrac{\eta m}{2\kappa}]$ and ${\cal C}_{\hat 2, \hat 3} = (-\infty , -\tfrac{2\kappa}{\eta m} ] \cup [ \tfrac{2\kappa}{\eta m} , \infty) $.

In the algebraic curve description, each cut ${\cal C}$ carries a  filling fraction ${\cal S}$, defined  as the A-cycle integral encircling ${\cal C}$ and giving a dimensionless measure of its length.
 On the spin chain side, this corresponds to the continuum filling of Bethe roots. A macroscopic (infinite) cut is then one whose filling fraction diverges and thus represents a classical condensate of Bethe roots, mapping to a classical sigma-model solution with non-vanishing charges. 
Microscopic (infinitesimal) cuts, on the other hand, correspond to one-magnon excitations on top of that condensate, and are understood to map to semiclassical quantum fluctuations of the sigma-model solutions. See e.g.~\cite{Kazakov:2004qf,Beisert:2005bm,Schafer-Nameki:2010qho,Gromov:2009zza} (and references therein).

In this picture, the BMN-like condensate with $m\neq 0$ is thus carried by the  macroscopic  ${\cal C}_{\hat 2, \hat 3}$ cut lying in the fundamental physical region, 
while ${\cal C}_{\hat 1, \hat 4}$ is its mirror, lying in $\lvert u \rvert \leq 1$, and has  finite filling fraction. 
In fact, as we will soon see,  the contribution from ${\cal C}_{\hat 2, \hat 3}$ dominates the vacuum energy in the large $J$ semiclassical regime, while the finite ${\cal C}_{\hat 1, \hat 4}$ does not contribute  until ${\cal O}(J^{-2})$.
Consequently,   the semiclassical  fluctuations above the BMN-like vacuum that must map to  the one-magnon excitations of the twisted $\mathrm{XXX}_{-1/2}$ spin chain at large $J$ should indeed be realised as microscopic cuts connecting the $\hat2$-$\hat3$ sheets, as mentioned earlier. In other words, the semiclassical $\mathfrak{sl}(2,R)$ sector is realised by the curve restricted to $\hat2$-$\hat3$.

Let us now see this explicitly and subsequently match with the spin chain results of section \ref{s:all-J-spinchain} in the  $J\rightarrow \infty$ limit.

\subsection{Semiclassical spectrum and matching} \label{s:string-large-J}

In \cite{Borsato:2022drc}, the semiclassical one-loop spectrum of the BMN-like solution was computed from its algebraic curve \eqref{eq:BMN-like-curve}.
We can write  the total  spectrum as
\[
\mathbf{E}_{\tts{tot.}} =  \mathbf{E}_0 + \mathbf{E}_1 ,
\]
where   $\mathbf{E}_0$ is the  vacuum  energy of the twisted solution and $\mathbf{E}_1$ is the sum over the semiclassical fluctuation modes. The vacuum generally receives both a classical and a semiclassical one-loop contribution, 
$ \mathbf{E}_0 = \E + {\E}_{\tts{1-loop}}$, 
with the classical $\E =\sqrt{\lambda} {\cal E}$ given in \eqref{eq:on-shell-class-energy}-\eqref{eq:class-energy} and 
\[
{\E}_{\tts{1-loop}}= \frac{1}{2}\sum_{n\in\mathbb Z}\sum_{ij}(-)^{F_{ij}} \Omega_n^{ij} ,
\]
while the excited states contribute at one-loop as\footnote{This sum is often denoted  as $\delta\Delta$, as   in \cite{Borsato:2022drc}.}
\[
\mathbf{E}_1 =  \sum_{n\in \mathbb{Z}} \Omega_n^{ij} N_n^{ij} .
\]
Here $N_n^{ij}$ is the occupation number of an excitation connecting sheets $i$ and $j$ with integer mode number $n$, while $\Omega_n^{ij}$ are the corresponding fluctuation frequencies. In addition, $F_{ij}=0$ for bosonic excitations (connecting only AdS sheets $\hat{1} \cdots \hat{4}$ or sphere sheets $\tilde{1} \cdots \tilde{4}$) and $F_{ij}=1$ for fermionic excitations (connecting an AdS with a sphere sheet).

As for the undeformed models, the overlap between the twisted string and nearest-neigbour spin chain spectra would arise in the large $J$ regime, where the spin chain becomes infinitely long (the thermodynamic limit) and the string admits the semiclassical description. However, as in the undeformed case, the nearest-neighbor 
approximation breaks down beyond ${\cal O}(J^{-3})$, also known as the three-loop discrepancy, when long-range interactions and 
finite-size effects become important \cite{Beisert:2005cw}.
We might thus similarly expect here that the vacuum energy $\mathbf{E}_0$  of the string 
maps to the ground-state energy $E_0$~\eqref{eq:E-J-S=0-O10} of the spin chain, measured on top of the untwisted BMN/BPS vacuum $\Delta = J$, i.e.\footnote{We  use the symbol $\sim$ to denote mappings between the string and spin chain models. }~
\[ \label{eq:GS-str-sc}
\mathbf{E}_0 ~~~ \sim  ~~~ \Delta + E_0 ,
\]
 up to order ${\cal O}(J^{-2})$. Furthermore, as argued earlier, in the $\mathfrak{sl}(2,R)$ sector  the string fluctuation energy $\mathbf{E}_1 \rvert_{\hat 2 \hat{3}} $ from the $\hat2$-$\hat3$ sheets 
should reproduce the spectrum $\delta E$ of multi-magnon excitations of the twisted $\mathrm{XXX}_{-1/2}$ spin chain above its ground state, again up to ${\cal O}(J^{-2})$, i.e. 
\[ \label{eq:ES-str-sc}
\mathbf{E}_1 \rvert_{\hat 2 \hat{3}} = \sum_{n\in \mathbb{Z}} N_n^{\hat 2 \hat{3}} \Omega_n^{\hat 2 \hat{3}}  ~~~ \sim  ~~~ \delta E = \sum_{n\in \mathbb{Z}}  N_n (E_{1,n} - E_0) ,
\]
with $E_{1,n}$  the single-magnon energies given in \eqref{eq:sc-E1-allJ}. Here we further assume that, at this order, magnon–magnon interactions are suppressed, so that the total energy is simply the sum of single-magnon contributions with occupation numbers $N_n \sim  N_n^{\hat 2 \hat{3}}$.

The fluctuation frequencies $\Omega_n^{ij}$ were computed explicitly in \cite{Borsato:2022drc}, and in the large-$J$ expansion  they take the form\footnote{In \cite{Borsato:2022drc} different conventions on overall factors of the $r$-matrix and notations for the charges were used; the simplest way to match parameters is at the level of the classical BMN-like solution. The relevant mapping here is $a_T = - \kappa$ and $\beta = \tfrac{\eta m}{2 \kappa}$.  }
\begin{equation} \label{eq:all-frequencies-expansion} 
\begin{aligned}    
    &\Omega^{\tilde{1} \tilde{3}}_n = \Omega^{\tilde{1} \tilde{4}} (x_n) = \Omega^{\tilde{2} \tilde{3}}_n =\Omega^{\tilde{2} \tilde{4}} (x_n) = \frac{\lambda n^2 }{2J^2} + {\cal O}(J^{-3})    \ , \\
    &\Omega^{\hat{1} \hat{4}}_n= \frac{ \lambda n^2}{2J^2} + {\cal O}(J^{-3})  \ , \\
    &\Omega^{\hat{2} \hat{3}}_n=  \frac{\lambda\sqrt{n^2  (n^2  + \eta^2 m^2)}}{2J^2} + {\cal O}(J^{-3})  \ ,  \\
    &\Omega^{\hat{1} \hat{3}}_n=\Omega^{\hat{2} \hat{4}} (x_n)=  \frac{\lambda ( 4 n^2   + \eta^2 m^2 )}{8 J^2} + {\cal O}(J^{-3})  \ , \\
    &\Omega^{\hat{1} \tilde{3}}_n=\Omega^{\hat{1} \tilde{4}} (x_n)=\Omega^{\tilde{1} \hat{4}}_n=\Omega^{\tilde{2} \hat{4}} (x_n)= \frac{ \lambda n^2}{2J^2} + {\cal O}(J^{-3})   \ , \\
    &\Omega^{\hat{2} \tilde{3}}_n=\Omega^{\hat{2} \tilde{4}} (x_n)=\Omega^{\tilde{1} \hat{3}}_n=\Omega^{\tilde{2} \hat{3}} (x_n)=   \frac{\lambda (4 n^2   + \eta^2 m^2 ) }{8 J^2} + {\cal O}(J^{-3})   \ .
\end{aligned}
\end{equation}
It is noteworthy that all expressions are even in the deformation parameter $\eta$. After the $J$-expansion, one can then calculate the sum $\E_{\tts{1-loop}}$ simply by using zeta-function regularisation, similarly as was done in \cite{Ouyang:2017yko}. That is, we use
\[
\sum_{n\in \mathbb{Z}} \left( (n+q)^2 + p n \right) =  q^2 + \zeta (-2,1+q) + \zeta (-2 , 1-q) =0 , \qquad \forall p,q ,
\]
where $\zeta ( s, a) := \sum_{n=0}^\infty \frac{1}{(n+a)^s}$. Hence, only the contribution from $\Omega^{\hat{2} \hat{3}}_n$ is nontrivial, which is consistent with ${\cal C}_{\hat 2, \hat 3}$ corresponding to the vacuum (macroscopic) cut, while all other fluctuations cancel in the (graded) sum. Notice that $\Omega^{\hat{2} \hat{3}}_n$ vanishes for $n=0$ and is even in $n$. For completeness, we can therefore write 
\[
\E_{\tts{1-loop}} &= \frac{1}{2} \sum_{n\in \mathbb{Z}} \Omega^{\hat{2} \hat{3}}_n =  \sum_{n=1}^\infty  \frac{\lambda \sqrt{n^2  (n^2  + \eta^2 m^2)}}{2J^2} + {\cal O}(J^{-3}) , \\
\mathbf{E}_1 \rvert_{\hat 2 \hat{3}} &=  \sum_{n\in \mathbb{Z}}  N_n  \frac{\lambda\sqrt{n^2  (n^2  + \eta^2 m^2)}}{2J^2}  + {\cal O}(J^{-3}) .
\]

Let us now finally compare to the spin chain results. Classically,   there is an explicit overlapping  regime known as the Landau-Lifshitz limit,  where the classical actions of the  spin chain and the sigma-model coincide at  $J\rightarrow\infty$ and fixed $\tilde{\lambda} = \lambda J^{-2}$ \cite{Kruczenski:2003gt,Stefanski:2004cw}. The only contribution to the spectrum at the string side comes from the classical vacuum energy
\[ \label{eq:str-class-vac-exp}
\E -J =  \frac{\tilde{\lambda} J \eta^2 m^2}{8} + \cdots =  \frac{\lambda \eta^2 m^2}{8J} + \cdots .
\]
Expanding $E_0$ from the spin chain \eqref{eq:E0-J-LL-match} when keeping $M$ and $\eta = \tfrac{\sqrt{\lambda}}{2\pi}\xi$ fixed\footnote{Fixing  $M$ and $ \eta$  in the classical worldsheet limit ensures that, at large $\lambda$ and fixed charges,  the sigma-model twist reproduces the classical limit of the spin-chain Drinfel'd twist under $\eta = \tfrac{\sqrt{\lambda}}{2\pi}\xi$, see e.g.~around eq.~(2.18) in \cite{Driezen:2025dww}. 
\label{f:map-twists-cl} } gives
\[
E_0 = \frac{J \tilde{\lambda}}{32\pi^2} \log (1+\xi M)^2 + \cdots = \frac{\eta^2 M^2}{8J} + \cdots ,
\]
as was also observed in \cite{Driezen:2025dww}. In the Landau-Lifshitz classical limit we thus have $\sqrt{\lambda} m \sim M$. 

Beyond this overlapping classical regime, a consistent mapping of quantum corrections is done by expanding at the string side
in powers of $1/\cJ$, with $\cJ = J / \sqrt{\lambda}$, and at the spin chain side in $1/J$ (see e.g.~\cite{Beisert:2005cw}). 
For $\E_{\tts{1-loop}}$ we get 
\[
\E_{\tts{1-loop}} &= \sum_{n=1}^\infty \frac{1}{\cJ^2} \left( \frac{n^2 }{2} + \frac{\eta^2 m^2}{4} - \frac{\eta^4 m^4}{16  n^2} + \frac{\eta^6 m^6}{32  n^4}  -\frac{5\eta^8 m^8}{256  n^6}  + {\cal O}(\eta^{10}, \cJ^{-2}) \right) + {\cal O}(\cJ^{-3}) , \\
&= \frac{1}{\cJ^2} \left(  \frac{\zeta(-2) }{2} + \frac{\zeta(0)\eta^2 m^2}{4} - \frac{\zeta(2)\eta^4 m^4}{16  } + \frac{\zeta(-4)\eta^6 m^6}{32  }  -\frac{5 \zeta(-6)\eta^8 m^8}{256  }  +  \cdots \right) \\
&= \frac{1}{\cJ^2} \left(  - \frac{\eta^2 m^2}{8} - \frac{\pi^2 \eta^4 m^4}{96 } + \frac{\pi^4\eta^6 m^6}{2880  }  -\frac{\pi^6 \eta^8 m^8}{48384  }    + {\cal O}(\eta^{10}, \cJ^{-2}) \right)  + {\cal O}(\cJ^{-3})
\]
where $\zeta(s) = \zeta(s,1)$ while $\E-J$  expands as in \eqref{eq:str-class-vac-exp}. Note that we have in addition expanded around $\eta=0$, given that the analytic Baxter results from section \ref{sec:Baxter} are similarly obtained only perturbatively in the deformation parameter. On the spin chain side, on the other hand,   we find it more natural  to expand in  the variable $\phi = \frac{i}{2} \log (1+\xi M)$ rather than $\eta$. This gives
\[ \label{eq:GS-E0-exp}
 E_0 =  -\frac{\lambda \phi^2}{8\pi^2 J} + \frac{\lambda}{J^2} \left( \frac{ \phi^2}{8\pi^2 } -\frac{ \phi^4}{96\pi^2 } - \frac{ \phi^6}{2880\pi^2 } - \frac{ \phi^8}{48384\pi^2 } +{\cal O}(\phi^{10}, J^{-2}) \right) + {\cal O}(J^{-3}) .
 %
 %
\]
We then see that upon the following identification of momentum charges
\[ \label{eq:Mstr-Mchain}
2\pi\eta m  ~~~\sim ~~~  \log (1+\xi M) , 
\]
there is a very nontrivial matching  \eqref{eq:GS-str-sc}  to ${\cal O} (J^{-2})$ of the one-loop ground state energy! 
Even more striking is that this agreement extends to the excited states; we have
\[
\mathbf{E}_1 \rvert_{\hat 2 \hat{3}} =  \sum_{n} N_n \frac{1}{\cJ^2} \left(    \frac{n^2}{2 } + \frac{\eta^2 m^2}{4 } - \frac{\eta^4 m^4}{16  n^2 } \right) + {\cal O}(\eta^6, \cJ^{-2}) + {\cal O}(\cJ^{-3}) , \\ 
\delta E = \sum_{n} N_n \frac{\lambda}{J^2} \left( \frac{ n^2 }{2} - \frac{ \phi^2}{4\pi^2 } - \frac{ \phi^4}{16\pi^4 n^2 } \right) + {\cal O}(\phi^6, J^{-2}) + {\cal O}(J^{-3}) ,
\]
which confirms \eqref{eq:ES-str-sc} to ${\cal O}(J^{-2})$ for any mode number $n$ under the same identification \eqref{eq:Mstr-Mchain}. 
This agreement is remarkable considering the highly complex form of the single-magnon energies $E_{1,n}$ given in \eqref{eq:sc-E1-allJ} up to ${\cal O}(\phi^4)$, where almost every numerical coefficient contributes to the subleading $J^{-2}$ correction. In fact,   it goes well beyond what one might expect from a simple 
lowest-order matching: it persists up to the same order  as in  undeformed AdS/CFT. Given the severely reduced  residual Noether (super)symmetry, one might    expect  the three-loop discrepancy induced by  long-range effects \cite{Beisert:2005cw} to appear much earlier here. 
That it does not suggests that the integrable structure itself, rather than the Noether algebra, provides the essential  mechanism stabilising the correspondence at this order.

Let us now discuss several important aspects of the matching we have found.

\begin{itemize}
\item \textit{Identifying momentum charges from matching the algebraic curve and the $Q$-functions.}\\
Let us  recall the classical relation \eqref{eq:Q-to-M} found from the spectral curve of the BMN-like string. In \cite{Borsato:2022drc} it was shown that $\mathbf{Q}$ does not receive any semiclassical corrections, implying that $\mathbf{Q} = 2\pi \eta  \sqrt{\lambda} m$ also holds semiclassically.  Together with
the identification \eqref{eq:Mstr-Mchain} we can thus conclude that the string twist charge $\mathbf{Q}$ relates  to the spin chain parameter through $\log (1+\xi M)$. This identification  is physically well-motivated: 
Firstly, at the classical level it reduces to  $\sqrt{\lambda} m \sim M$, as we found in the Landau-Lifshitz limit,  when identifying $\eta = \tfrac{\sqrt{\lambda}}{2\pi} \xi$ (exactly  as was needed when mapping the string and spin chain Drinfel'd twist, cf.~footnote \ref{f:map-twists-cl}).
At the quantum level, one can understand  the logarithmic structure  as reflecting the non-abelian nature of the Jordanian twist. 
In particular, this should be contrasted with   abelian twists such as the dipole/Schr\"odinger deformation  where the momentum charges map also at the quantum level simply as $\sqrt{\lambda} m \sim M$ \cite{Guica:2017mtd,Ouyang:2017yko}.

A posteriori, the identification \eqref{eq:Mstr-Mchain} is also consistent with the structural relationship 
between spectral curves and Baxter $Q$-functions. In fact, semiclassically,  the quasimomenta of the algebraic  curve are connected to the $Q$-functions through the so-called resolvent  $G(u)$ \cite{Gromov:2009zza}. In the undeformed case the latter is defined on the spin chain side as  $G(u) = \sum_n (u-u_n)^{-1}$ for a discrete set of 
Bethe roots $\{u_n\}$,  and $Q(u)$ is as in \eqref{eq:undef-Q}. On the string side,  the classical spectral curve likewise 
defines a resolvent through the quasimomenta as $G(u) = p(u) + \tfrac{p_+}{1+u} + \tfrac{p_-}{1-u}$, where the latter terms remove the pole structure from $p(u)$.\footnote{Note that this string relation is more general and not restricted to the undeformed case.} The undeformed correspondence thus identifies
\[ \label{eq:p-to-Q}
 p(u) + \frac{p_+}{1+u} + \frac{p_-}{1-u} ~~~\sim ~~~ \partial_u \log Q(u) .
\]

For the Jordanian-twisted case, neither the $Q$-functions nor the spectral curve  retain their undeformed  structure: the twisted  $\mathfrak{sl}(2,R)$  $Q$-functions  are non-polynomial with exponential asymptotics \eqref{eq:Q_asymptotics}, while the $\hat{p}_2(u)$ and $\hat{p}_3(u)$ quasimomenta  in the  $\mathfrak{sl}(2,R)$ sector  acquire a constant twist term \eqref{eq:large-u-Jor-p}. It is therefore not obvious whether the structural relation 
\eqref{eq:p-to-Q} between resolvents, $Q$-functions, and quasimomenta should persist. 
Nevertheless, if one assumes this relation continues to hold,  then taking its large-$u$ 
asymptotics reproduces precisely 
\[
\frac{\mathbf{Q}}{\sqrt{\lambda}} ~~~\sim ~~~ \log (1+\xi M) ,
\]
up to an ambiguous assignment of signs. 
The fact that this assumption correctly reproduces the nontrivial spectral matching \eqref{eq:GS-str-sc} and \eqref{eq:ES-str-sc} up to $O(J^{-2})$ provides strong evidence that the   relation \eqref{eq:p-to-Q} indeed persists as a structural feature of the integrable  correspondence between strings and spins, even under non-abelian deformations that drastically modify the 
analytic properties of both $Q$-functions and quasimomenta.

\item \textit{Ordering of $J$-expansion and resummation.}\\
In \cite{Borsato:2022drc}, the  one-loop correction $\E_{\tts{1-loop}}$ was obtained by converting the discrete mode sum to an integral under the assumption of a large energy density $\kappa = \tfrac{\E}{\sqrt{\lambda}}$. Expanding their re-summed result around large $J$ then gives, in our conventions, 
\[
\E_{\tts{1-loop}} = \frac{\eta^3 \M^3}{6 \sqrt{\lambda} J^2} + {\cal O}(J^{-3}) ,
\]
which does not coincide with the ${\cal O}(J^{-2})$ term of \eqref{eq:GS-E0-exp}. 
By contrast, our computation expands the fluctuation frequencies at large $J$ prior to resummation (performed via zeta-function regularisation). These two limits need not commute, given that $\kappa$ need not be large in the worldsheet semiclassical regime, which may  alter results. 
The matching found here suggests that the large-$J$ expansion prior to resummation provides the physically appropriate regularisation scheme for the comparison. It would nevertheless be valuable to clarify this order-of-limits ambiguity by explicitly analysing possible $\kappa^{-1}$ corrections in the procedure of \cite{Borsato:2022drc}. 

 A related subtlety concerns the inclusion of fermionic (super-)modes. Performing the resummation first, as in \cite{Borsato:2022drc}, implicitly incorporates the complete set of fluctuations, while our large-$J$ expansion effectively isolates the bosonic $\mathfrak{sl}(2,R)$ sector prior to supersymmetric completion. This distinction may be meaningful, as  the bosonic Jordanian $r$-matrix by itself does not yield a genuine type IIB supergravity background, which only emerges once the $r$-matrix is extended by supercharges \cite{2004Tolstoy,vanTongeren:2019dlq,Borsato:2022ubq}. 
  It is thus also possible that the apparent sensitivity to the order of expansion and resummation  reflects this difference between the purely bosonic and complete realisation of the deformation within the AdS/CFT context, which could be clarified by extending the comparison to the full super-spin chain framework.
\end{itemize}

\section{Conclusions} \label{sec:conclusions}

In this work, we have established the solvability 
of the Jordanian-twisted $\mathrm{XXX}_{-1/2}$ spin chain through the Baxter framework, extending previous results  to higher excitations and arbitrary spin chain length, and providing compelling evidence for a non-abelian, Jordanian-twisted, AdS/CFT correspondence.   
Building on \cite{Driezen:2025dww}, we attacked this from two complementary approaches: direct diagonalisation of the $J=2$     nearest-neighbour Hamiltonian (perturbative in the deformation parameter) and  the Baxter $TQ$-relation.  As a result we can efficiently compute the full deformed spectrum of this spin chain.

We proposed a Baxter framework for the  Jordanian deformation with intriguing  and fully novel structural features   compared to previously known cases in the literature (as summarised in section \ref{sec:qas}). 
In particular, despite the fact that the  Yangian (RTT) structure is deformed,  we propose that the functional form of the $TQ$-relation and its energy formula remain \textit{undeformed}. 
The deformation enters only through fixed coefficients in the transfer matrix, resulting in non-polynomial $Q$-functions with exponential asymptotics.
An additional crucial ingredient is that analytic regularity in the complex plane replaces the polynomiality requirement of undeformed chains in order to select   spectral $Q$-functions, similar as in \cite{Guica:2017mtd}. 
This proposal 
is supported by several highly non-trivial tests. The direct computation of the $J=2$ spectrum from the Hamiltonian  provides a first crucial validation, as it perfectly matches the spectral results from Baxter across higher spin sectors. 
We also show more generally that imposing regularity of the $Q$-function fixes the transfer matrix eigenvalue (and the $Q$-function itself). In fact it is non-trivial that a regular solution for the $Q$-function exists at all,  further providing strong evidence in favour of our proposal. 
Importantly, this also supports the expectation that the functional form of the  $TQ$-relation is universal for all models based on rational $R$-matrices, as commented on in \cite{Guica:2017mtd,Driezen:2025dww}, even when the underlying algebraic structure is substantially modified.\footnote{Another potentially useful argument is the observation that Drinfel'd twists reduce to the identity in one-dimensional representations. This fact is relevant because the form of the $TQ$ relation (its $(u\pm i/2)^J$ coefficients) is expected to be fixed by transfer matrices with one-dimensional auxiliary spaces (see  \cite{Bazhanov:2010ts} or \cite{Frenkel:2013uda}), in which the Drinfel'd twist would never appear, even for a Drinfel'd twisted $R$-matrix/$RTT$ algebra.   }
In the future it would be highly interesting to rigorously derive the Baxter equation and the regularity condition from a full operatorial construction of $Q$-functions, e.g.~along the lines of \cite{Chervov:2006xk, Bazhanov:1996dr,Bazhanov:2010ts} or \cite{Derkachov:2001yn,Derkachov:2002tf}.

The compatibility between the analyticity conditions for the $Q$-functions of the Jordanian and dipole deformation \cite{Guica:2017mtd} led us further to consider  a more general model, combining in spirit both deformations (yet whose physical realisation we leave open). We illustrated for $J=2$ and $S=0$ that its spectrum can also be solved with the Baxter equation.  

With the Baxter framework  provided for the Jordanian chain, we efficiently computed the  spectrum to arbitrary length $J$ for the ground and first excited states. This enabled comparison with the 
string worldsheet theory, 
conjectured via the work of \cite{vanTongeren:2015uha} to be the Jordanian HYB-deformation in the $\mathfrak{sl}(2,R)$ sector. In particular,  the spectrum of the (nearest-neighbour) twisted chain in the large-$J$ limit matches rather fascinatingly  with the 
semiclassical string spectrum  \cite{Borsato:2022drc} from the  algebraic curve to order ${\cal O}(J^{-2})$ included---exactly the same order at which the correspondence holds as in undeformed AdS/CFT  before long-range interactions trigger discrepancies \cite{Beisert:2005cw}. This is particularly striking given that    the Jordanian model  has very little Noether  (super)symmetry. The agreement despite this symmetry reduction suggests that integrability itself appears sufficient to stabilise the correspondence at this order. 

The spectral agreement furthermore required a particular identification of the twist charges $\tfrac{\mathbf{Q}}{\sqrt{\lambda}} \sim  \log (1+\xi M) $, which
we discussed in detail at the end of section \ref{sec:strings}. Importantly, this identification is supported from matching the asymptotics of both the algebraic curve and $Q$-functions, 
providing a compelling consistency check of the entire framework.

Looking forward, several important directions remain open: 

\noindent \textbf{Hilbert space.}  While energy eigenvalues deform smoothly from their undeformed values, understanding the underlying Hilbert space structure remains subtle. 
In particular, explicit mapping of  the $\mathfrak{sl}(2,R)$ highest-weight module to eigenstates of its  root generator would help clarify  relations to factorised worldsheet scattering \cite{Borsato:2024sru} and the generalised eigenstates discussed in \cite{Borsato:2025smn}. 

\noindent \textbf{Separation of Variables (SoV).} Our results provide the foundation for a full SoV treatment. In recent years, much attention has been brought to this programme (notably in the context of AdS/CFT correlators) with numerous novel developments   for new classes of spin chains as well as field theories (see e.g.~\cite{Maillet:2018bim, Cavaglia:2019pow,Gromov:2019wmz} and the review \cite{Levkovich-Maslyuk:2025ipl}). Given the intriguing features of the Jordanian chain, performing the full SoV programme would be valuable.
In particular,  an operatorial construction of the $TQ$-relation and determination of the separated variable bases would yield additional structural insights, including on the structure of the Hilbert space.

\noindent \textbf{Beyond semiclassical matching.} The ${\cal O}(J^{-2})$ agreement with string theory opens several exciting directions: \textit{(i)} formulating the full Quantum Spectral Curve (proposed for  undeformed ${\mathcal N}=4$ SYM in \cite{Gromov:2013pga}) in order to extend the correspondence beyond ${\cal O}(J^{-2})$  and one-loop, thereby incorporating finite-size effects. This is particularly promising given  the relatively simple asymptotic   behaviour of the $Q$-functions as well as the undeformed functional forms of the Baxter $TQ$-relation and the  identification of string quasimomenta with $Q$-functions; 
\textit{(ii)} understanding the robustness of the twist identification $\tfrac{\mathbf{Q}}{\sqrt{\lambda}} \sim  \log (1+\xi M) $  under  finite-size corrections; and 
\textit{(iii)} extending the analysis to the full $\mathfrak{psu}(2,2|4)$ superchain, which would require including fermionic modes and unimodular contributions to the Jordanian (Drinfel'd) twist.

\noindent \textbf{Gauge theory construction.} 
The successful one-loop matching between the string and spin chain spectrum strongly motivates constructing a Jordanian deformation of ${\cal N}=4$ SYM. Since the Jordanian  
$r$-matrix  
affects the $\mathfrak{so}(2,4)$ isometries, the gauge theory will have a    non-commutative spacetime. 
Recent  progress   in such non-commutative  deformations  \cite{Meier:2023kzt,Meier:2023lku,Meier:2025tjq} offers a useful foundation. It would   enable field-theoretic computations of observables, important for non-AdS holography in general, and provide a complete picture of Jordanian AdS/CFT across all three corners of the correspondence. 

\section*{Acknowledgements}

We thank N.~Beisert, S.~Ekhammar, G.~Felder, G.~Ferrando, N.~Gromov, E.~Im, D.~Serban, C.~Thull, A.~Tseytlin, B.~Vicedo and K.~Zarembo for discussions, and K.~Zarembo for comments on the draft. The work of F.L.-M. is supported by the STFC grant APP69281. The work of S.D.~is supported by the Swiss National
Science Foundation through the SPF fellowship TMPFP2$\_224600$ and both A.M.~and S.D.~acknowledge support by the NCCR SwissMAP.

\appendix
\section{Closed-form  $J=2$ eigenstates for $S=0,1$}\label{app:J=2_eigenstates}
For deformations of the lowest excitations $S=0,1$, and upon appropriate normalisation as explained in the main text,   one can directly solve the ODEs in the lowest orders in $\xi$ without having to perform a Frobenius analysis. This thus gives compact analytic expressions for   the $J=2$ eigenfunctions   in the relative coordinate $z$. 
For $S=0$, they read
\[
g_{S=0}&(z)=\frac{2 \sinh \left(\frac{M z}{2}\right)}{M z}+\xi  \left(\frac{\cosh \left(\frac{M z}{2}\right)}{2 z}-\frac{\sinh \left(\frac{M z}{2}\right)}{M z^2}\right)\\
    &+\xi ^2 \left(\frac{(M z (2 M z+9)+24) \sinh \left(\frac{M z}{2}\right)}{18 M z^3}-\frac{(3
   M z+8) \cosh \left(\frac{M z}{2}\right)}{12 z^2}\right) +{\cal O}(\xi^3 M^3),
\]
which is exact in $z$ up to order  ${\cal O}(\xi^3 M^3)$ while
for $S=1$, we find
\[
g_{S=1}&(z)=\mathrm{SJ}\left(1;\frac{i M z}{2}\right)-\xi  \Bigg(-\frac{8 i \mathrm{SJ(-2;}\frac{i M z}{2}) \sinh (M z)}{M^2 z^3}+\frac{8 \mathrm{SY(-2;}\frac{i M z}{2}) \cosh (M z)}{M^2 z^3}\\
        &+\frac{i \left(3 M^2 z^2+16\right) \mathrm{SJ(-2;}\frac{i M
   z}{2}) \cosh (M z)}{2 M^3 z^4}-\frac{\left(3 M^2 z^2+16\right) \mathrm{SY(-2;}\frac{i M z}{2}) \sinh (M z)}{2 M^3 z^4}\\
        &+\frac{i \left(M^4 z^4+10 M^2 z^2-32\right) \mathrm{SJ(-2;}\frac{i M
   z}{2})}{4 M^3 z^4}-\frac{3 \mathrm{SY(-2;}\frac{i M z}{2})}{2 z}\Bigg) +{\cal O}(\xi^2 M^2),
\]
with $\mathrm{SJ}$ and $\mathrm{SY}$ the spherical Bessel function of the first and second kind respectively, and exact in $z$ up to order ${\cal O}(\xi^2 M^2)$.

\bibliographystyle{JHEP.bst}
 \bibliography{jordbib}

\end{document}